\documentclass[prd,showpacs,preprint,preprintnumbers,amsmath,amssymb,eqsecnum,nofootinbib]{revtex4-1}
\usepackage{graphicx,amssymb,amsmath,amsthm,amsfonts,mathrsfs,epsfig,epsf}
\usepackage{dcolumn}
\usepackage{amsfonts,mathrsfs}
\usepackage{bm}
\usepackage{subfigure}
\usepackage{color}
\usepackage{url}

\newcommand{\eq}[1]{\mbox{Eq.~(\ref{#1})}}
\newcommand{\fig}[1]{\mbox{Fig.~\ref{#1}}}
\newcommand{\D}{{\rm d}}

\begin{document}

\begin{titlepage}
\vfill

\vfill
\begin{center}
\baselineskip=16pt
{\Large\bf  Thin-shell wormholes in Einstein and Einstein-Gauss-Bonnet theories of gravity
}
\vskip 0.5cm
{\large {\sl }}
\vskip 10.mm
{\bf  Takafumi Kokubu${}^{a}$ 
 and Tomohiro Harada${}^{b}$} \\

\vskip 1cm
{
    ${}^a$ Department of Physics and Synergetic Innovation Center for Quantum Effects and Applications, Hunan Normal University, Changsha, Hunan, China\\
 ${}^{b}$ Department of Physics, Rikkyo University, Toshima, Tokyo, Japan
     }
\vspace{6pt}
\end{center}
\vskip 0.2in
\par
\begin{center}
{\bf Abstract}
 \end{center}
\begin{quote}

We review recent works on the possibility for eternal existence of thin-shell wormholes on Einstein and Einstein-Gauss-Bonnet gravity. We introduce thin-shell wormholes that are categorized into a class of traversable wormhole solutions. After that, we discuss stable thin-shell wormholes with negative-tension branes in Reissner-Nordstr\"om-(anti) de Sitter spacetimes in $d$ dimensional Einstein gravity. Imposing $Z_2$ symmetry, we construct and classify traversable static thin-shell wormholes in spherical, planar and hyperbolic symmetries. It is found that the spherical wormholes are stable against spherically symmetric perturbations. It is also found that some classes of wormholes in planar and hyperbolic symmetries with a negative cosmological constant are stable against perturbations preserving symmetries. In most cases, stable wormholes are found with the appropriate combination of an electric charge and a negative cosmological constant. However, as special cases, there are stable wormholes even with a vanishing cosmological constant in spherical symmetry and with a vanishing electric charge in hyperbolic symmetry. Subsequently, the existence and dynamical stability of traversable thin-shell wormholes with electrically neutral negative-tension branes is discussed in Einstein-Gauss-Bonnet theory of gravitation. We consider radial perturbations against the shell for the solutions, which have the $Z_2$ symmetry. The effect of the Gauss-Bonnet term on the stability depends on the spacetime symmetry.

  \vfill
\vskip 2.mm
\end{quote}
\thispagestyle{empty}
\end{titlepage}



\section{Introduction}\label{Introduction}
Wormholes are spacetime structures which connect two different universes or even two points of one universe. Wormholes have fascinated people for a long time and  many sci-fi movies and novels  are based on them. One may be surprised to know that wormholes are indeed a subject of theoretical physics. 
Unlike black holes, wormholes are hypothetical objects in a real world. Still, theoretical physicists have pursued such peculiar objects and revealed many properties of them because the consequences of wormhole physics are appealing; they enable instant travels between two distinct points and even realization of time travel \cite{mty1988}. 

To understand what a wormhole is, it is better to follow the history of wormholes first.
Interestingly, an insight into a wormhole shares the same year of discovery as the first black hole. In 1916  Schwarzschild found his famous black hole solution of the Einstein equations, the Schwarzschild black hole.
In 1916, Flamm found that the Schwarzschild metric has a hidden tunnel structure which connects two asymptotically flat spaces; he developed what is now called the embedding diagram \cite{Flamm}. 
 Almost twenty years later after Flamm's work, Einstein and Rosen published their famous paper about the ``Einstein-Rosen bridge'' which is the Schwarzschild spacetime that can be interpreted as a solution joining the two same Schwarzschild geometries at their horizons \cite{einstein-rosen1935}. The bridge acts like a spacetime tunnel since it connects two asymptotically flat regions.

The outline of this article is following:
Section \ref{Introduction} is the introduction, where
wormhole properties are reviewed.
In section \ref{Thin-shell wormhole}, we show the construction and linear stability against perturbations preserving symmetries of thin-shell wormholes.
In section \ref{Generalized Thin-shell wormholes}, we mention that there are many generalizations of the thin-shell wormhole by Poisson and Visser. In this section we review some of its generalizations.
In section \ref{Pure tension wormholes in Einstein gravity}, we concentrate on pure tension wormholes in Einstein gravity. This part is based on Ref. \cite{kh2015}.
We introduce wormholes with a negative-tension brane and we analyze the existence of static solutions,
stability and horizon avoidance in spherical, planar and hyperbolic symmetries. 
In section \ref{Pure tension wormholes in Einstein-Gauss-Bonnet gravity}, we treat pure tension wormholes in Einstein-Gauss-Bonnet gravity. This part is based on Ref. \cite{kokubu-maeda-harada2015}.
At first, assuming that the shell is made of tension together with a
perfect fluid, we derive the equation of motion for the shell and review
the basic properties of the static shell 
After that, we show that the shell has a negative energy density and
hence the weak energy condition is violated. (However, the negative-tension brane still satisfies the null energy condition).
Next, we study the existence and instability of static spherically symmetric thin-shell wormholes. 
The signature of the 
metric is taken as diag$(-,+,+,\cdots,+,+)$, and Greek indices run over all spacetime indices. In this section the $d$-dimensional gravitational constant $G_d$ is retained.
Section \ref{Discussions and conclusions} is dedicated to conclusion.

\subsection{Einstein-Rosen bridge} \label{sec-ER-bridge}
The Einstein-Rosen bridge is unstable since the throat pinches off quickly. To understand this mechanism, let us see the dynamics of the bridge to understand the reason of the pinch off.

\begin{figure}[htbp]
\begin{center}
\includegraphics[width=0.7\linewidth]{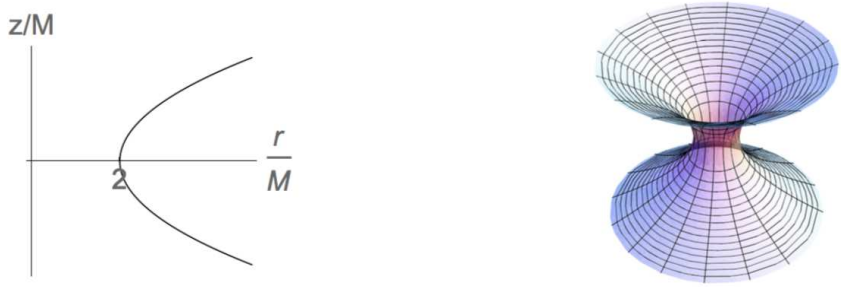}
   \caption{Left: The plot of \eq{z=z(r)}. Right: The surface is obtained by rotating the function around $z$ axis in the $\phi$ direction. The narrowest surface $z=0$ corresponds to $r=2M$.}
   \label{fig-embedding}
\end{center}
\end{figure}

We will show here how this tunnel structure is recognized. 
The Schwarzschild metric is written in the spherical coordinates $(t,r,\theta,\phi)$ as
\begin{align}
\D s^2=-\left(1-\frac{2M}{r}\right)\D t^2+\left(1-\frac{2M}{r}\right)^{-1}\D r^2+r^2(\D \theta^2+\sin^2 \theta \D \phi^2),
\end{align}
where $M$ is a constant mass parameter. Suppose we take a constant time slice, $t=$const. Since the spacetime is spherically symmetric, we can take the equator slice, $\theta=\pi/2$, without loss of generality. Then the metric reduces to
\begin{align}
\D s^2|_{t={\rm const.}, \theta=\pi/2}=\left(1-\frac{2M}{r}\right)^{-1}\D r^2+r^2 \D \phi^2. \label{equator-metric}
\end{align}
Eq. (\ref{equator-metric}) has an axial symmetry: $\phi \rightarrow \phi+$const.
The metric or the line element can be expressed as a two-dimensional surface in a three-dimensional flat space;
$ds^2$ at $t=$const and $\theta=\pi/2$ is embedded into the three dimensional Euclidean space $\mathbb{R}^3$.
In $\mathbb{R}^3$, the infinitesimal distance $\D \Sigma^2$ with the cylindrical coordinates ($\rho, \psi, z$) ($\rho$: distance from the $z$ axis, $\psi$: angle around the $z$ axis) is given by
\begin{align}
\D \Sigma^2=\D \rho^2+\rho^2 \D \psi^2+\D z^2. \label{cylinder-metric}
\end{align}
A surface in a flat space can be described as $z=z(\rho,\psi)$ in the cylindrical coordinates. Since the coordinates $r$ and $\phi$ are functions of $\rho$ and $\psi$, we must have relation between ($\rho, \psi, z$) and ($r, \phi$) to identify the equation of the surface, i.e., $\rho=\rho(r,\phi), ~~\psi=\psi(r,\phi),~~ z=z(r,\phi)$.
Since the metric (\ref{equator-metric}) is axially symmetric which suggests $\psi=\phi$ and $\rho=\rho(r)$, the surface function becomes the function of $r$, $z=z(r)$. Summarizing the above, we find
\begin{align}
\rho=\rho(r),~~\psi=\phi, ~~z=z(r). \label{rho-phi-z}
\end{align}
Substituting \eq{rho-phi-z} into \eq{cylinder-metric}, then comparing this metric and \eq{equator-metric}, we get the relations
\begin{align}
\left(\frac{\D z}{\D r}\right)^2+\left(\frac{\D \rho}{\D r}\right)^2=\left(1-\frac{2M}{r}\right)^{-1},~~\rho^2=r^2.
\end{align}
The simultaneous equations reduces to a single differential equation and is easy to integrate;
\begin{align}
\left(\frac{\D z}{\D r}\right)^2=\frac{2M}{r-2M}~ \Rightarrow ~z=\pm2\sqrt{2M(r-2M)}. \label{z=z(r)}
\end{align}
A plot for \eq{z=z(r)} is shown in \fig{fig-embedding} below in the assumption $M>0$. One finds that two asymptotically flat regions ($\D z/\D r\to0$ as $r\to\infty$) are connected by the neck  $z=0$. Due to the shape of a neck, we call it a {\it throat}.
%
At this stage, one may ask a question such as ``if we live in an asymptotically flat region, namely, in the upper space of \fig{fig-embedding}, what does the other region correspond to? Can we pass through the throat and go to this other region?". A clear answer to this question, is produced in the paper by Fuller and Wheeler \cite{fuller-wheeler1962}. They revealed the dynamical properties of the throat and also showed that no traveller can safely pass through the throat to go to the other region. We will show the dynamics of the throat by the following this argument.

\begin{figure}[htbp]
\begin{center}
\includegraphics[width=0.35\linewidth]{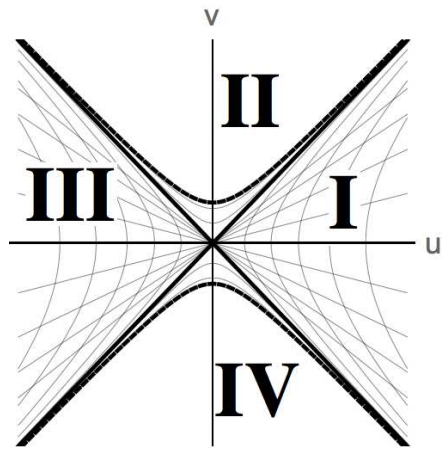}
   \caption{Kruskal diagram.}
   \label{fig-kruskal}
\end{center}
\end{figure}

\begin{figure}[htbp]
\begin{center}
\includegraphics[width=0.45\linewidth]{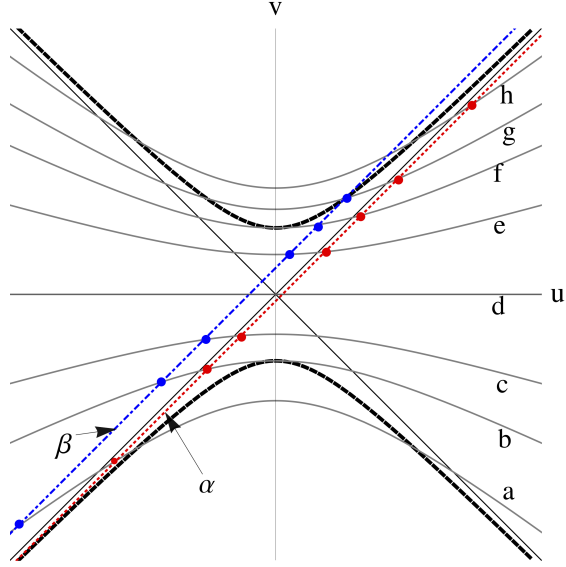}
\caption{The red dotted line is the geodesic of photon $\alpha$ from the region IV, while the blue one is that of photon $\beta$ from the region III. After passing through the anti horizon $v=-u$, the photon $\alpha$ goes to the region I and never across the event horizon $v=u$. the geodesic of the photon $\beta$ must terminate at the singularity $r=0$ in a finite time in the region II. }
   \label{fig-dynamics-kruskal}
\end{center}
\end{figure}

\begin{figure}[htbp]
\begin{center}
\includegraphics[width=0.7\linewidth]{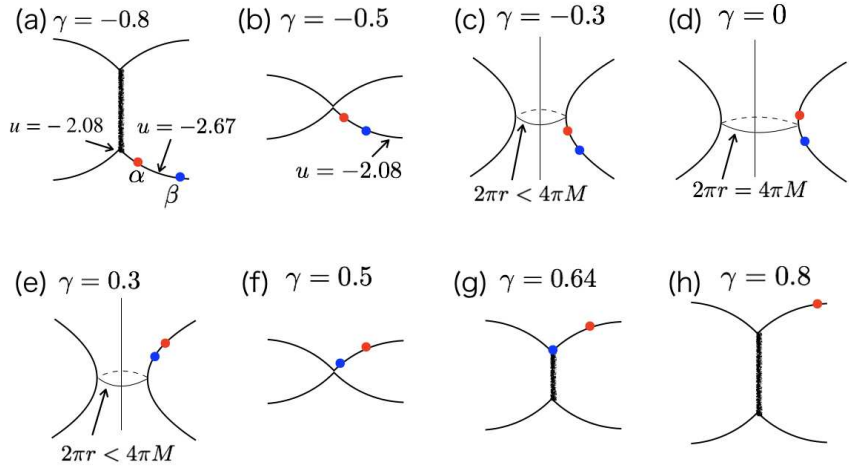}
   \caption{The dynamics of the throat and the motion of photons. A throat emerges instantaneously and connects two asymptotically flat space-sheets. After that, it expands and then starts to contract. Finally, it pinches off the connection between the two space-sheets.}
   \label{fig-throat-dynamics}
\end{center}
\end{figure}

The Kruskal diagram of the Schwarzschild spacetime is given by \fig{fig-kruskal}. Here, $v$ is a timelike coordinate while $u$ is a spacelike coordinate. Region I and III represent the outside of the black hole which corresponds to the region of $r>2M$ in the Schwarzschild coordinates. Region II is the inside of the black hole, $r<2M$, while region IV is the white hole, that is considered as a time-reversal of the black hole. The straight lines $v=\pm u$ correspond to $r=2M$, the event horizon of the spacetime. The dashed bold lines are the curvature singularity $(r=0)$. Straight lines between $v=+u$ and $v=-u$ are $t=$const., while hyperbolas are $r=$const. surfaces.
Here, we take a particular spacelike slice for the diagram as
\begin{align}
\gamma=\frac{v}{\sqrt{4+u^2}}, ~~\gamma={\rm const}. \label{space-slice}
\end{align}
which becomes $v=u$, as $u\to \pm \infty$. We draw $\gamma=$const. surfaces of \eq{space-slice}  in \fig{fig-dynamics-kruskal} as gray curves ((a) to (h)). As one can see, $\gamma$ plays the role of time here; when $\gamma$ increases, the surface \eq{space-slice} moves in the direction of increasing $v$. Since the surface \eq{space-slice} is spacelike, it moves in the time direction. 
In this figure, the red dotted straight line describes a null geodesic $\alpha$ released from the region IV, while the blue one is a null geodesic $\beta$ released from region III.  The throat cannot stay static and its dynamics is as described in \fig{fig-throat-dynamics}. The process occurs in the order of (a) to (h);\\
(a) Photons $\alpha$ and $\beta$ initially are in the lower sheet. They go to the center $r=0$. The values $u=-2.67$ and $u=-2.08$ correspond to $r=2M$ and $r=0$, respectively. The vertical bold line is the curvature singularity $r=0$. At this moment, the singularity is in between two quasi Euclidean spaces.\\
(b) Both photons go to the center. A throat is going to appear.\\
(c) The throat just opened. The circumference of the throat is smaller than $4\pi M$.\\
(d) The maximal throat, $2\pi r=4\pi M$. The photon $\alpha$ has passed though the throat.\\
(e) The throat is shrinking. Both of photons have passed though the throat.\\
(f) The moment of the throat closing. In this stage, both photons are still in the upper sheet, while the photon $\beta$ approaches the central singularity.\\
(g) Photon $\beta$ is just caught. Then, $\beta$ disappears in the singularity and stops existing.\\
(h) Photon $\alpha$ keeps escaping to the null infinity of the upper sheet.\\
Although we have used a specific spacelike slice \eq{space-slice} to show the dynamical feature of the bridge, this dynamics does not change as long as the slice is spacelike.

\begin{figure}[htbp]
\begin{center}
\includegraphics[width=0.3\linewidth]{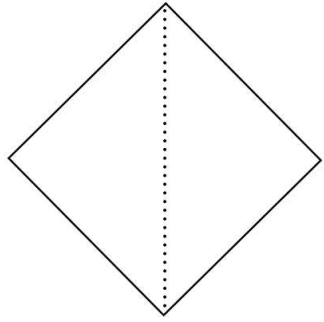}
   \caption{A Penrose diagram for a two-way traversable spacetime. The center vertical line describes a wormhole throat which connects the left and the right regions. From the above picture, a timelike traveller clearly can pass through the throat to go to the other region and also come back to the original region.}
   \label{fig-traversable-wormhole}
\end{center}
\end{figure}

From the above, we conclude that although a timelike traveller might go to the upper space in just a finite time but cannot come back to the lower space. Hence, the Schwarzschild solution provides a one-way travel.
One can speculate that a two-way travel needs a Penrose diagram like \fig{fig-traversable-wormhole}.  Apparently, this diagram shows that a timelike worldline can cross the throat again and again without hitting any singularities. The introduction of such a two-way tunnel spacetime is explained in the next subsection. 

\newpage
\subsection{Wormhole properties in brief} \label{Wormhole properties in brief}
It was Morris and Thorne who established modern wormhole physics. In this subsection, we follow their approach to understand what properties wormholes should have. 
They have pioneered qualitative study for static and spherically symmetric spacetimes which have "two-way" {\it traversable} wormholes \cite{morris-thorne1988}.
Since they knew what kinds of geometries describe tunnel structures, they deduced the metric which has such a geometry. Then, substituting the metric into the Einstein equation, they recovered the matter properties and its distribution. Here we begin with a brief overview of their discussion.

A convenient choice of coordinates to describe static and spherically symmetric wormhole spacetimes is
\begin{align}
\D s^2=-e^{2\Phi(r)}\D t^2+\left(1-\frac{b(r)}{r} \right)^{-1}\D r^2+r^2(\D \theta^2+\sin^2 \theta\D \phi^2), \label{morris-thorne-metric}
\end{align}
where $\Phi$ and $b$ are both functions of $r$.
To simplify calculations, we introduce an orthonormal basis of reference frame of the static observers:
\begin{align}
e^\mu_{\hat t}\partial_\mu=e^{-\Phi}\partial_t, ~~
e^\mu_{\hat r}\partial_\mu=\sqrt{1-\frac{b}{r}}\partial_r, ~~
e^\mu_{\hat \theta}\partial_\mu=\frac{1}{r}\partial_\theta, ~~
e^\mu_{\hat \phi}\partial_\mu=\frac{1}{r\sin \theta}\partial_\phi.
\end{align}
In this basis, the metric takes the Minkowskian form : $g_{\hat \mu \hat \nu}=g_{\alpha\beta}e^\alpha_{\hat \mu}e^\beta_{\hat \nu}=\eta_{\mu \nu}$. Then the non-zero components of the Einstein tensor yield
\begin{align}
&G_{\hat t\hat t}=\frac{r'}{b^2}, ~~
G_{\hat r\hat r}=2\left(1-\frac{b}{r} \right)\frac{\Phi'}{b^2}, \\
&G_{\hat \theta \hat \theta}=G_{\hat \phi \hat \phi}=\left(1-\frac{b}{r} \right)\left(\Phi''-\Phi'\frac{b'r-b}{2r(r-b)}+(\Phi')^2+\frac{\Phi'}{r}-\frac{b'r-b}{2r^2(r-b)}  \right), 
\end{align}
where $':=\D/\D r$. 
Since the geometry is both static and spherically symmetric, the vacuum equation must be the Schwarzschild black hole (Birkhoff's theorem), a non-traversable wormhole. Thus, if we want to build a wormhole spacetime we must handle spacetimes with a specific form of stress-energy tensors. As the Einstein tensor takes a diagonal form, the corresponding non-zero stress-energy tensor must also be diagonal. In the orthonormal basis we can then give each component of the stress-energy tensor the physical interpretation as
$T_{\hat t\hat t}=\rho(r),~T_{\hat r\hat r}=-\tau(r),~T_{\hat \theta \hat \theta}=T_{\hat \phi \hat \phi}=p(r)$,
where $\rho$ is the energy density, that static observers measure, $\tau$ is the radial tension that they measure in the radial direction (negative of the radial pressure), and $p$ is the pressure that they measure in the lateral direction.

The Einstein equations $G_{\hat \mu \hat \nu}=8\pi T_{\hat \mu \hat \nu}$ give the following non-trivial equations for $\rho, \tau$ and $p$:
\begin{align}
\rho=\frac{b'}{8\pi r^2},~~
\tau=\frac{1}{8\pi r^2}\left(\frac{b}{r}-2(r-b)\Phi' \right),~~
p=\frac{r}{2}\left((\rho-\tau)\Phi'-\tau' \right)-\tau. \label{morris-thorne-p}
\end{align}
One may solve the above equations to get the form of $b$ and $\Phi$ by imposing a specific component choice for $T_{\hat \mu \hat \nu}$, i.e., a specific form of $\rho, \tau$ and $p$. An alternative way to solve them is that one imposes an equation of state as $\tau=\tau(\rho)$ and $p=p(\rho)$, and then one solves for \eq{morris-thorne-p}.

\subsubsection{Embedding wormholes and asymptotic flatness}\label{Embedding wormholes and asymptotically flatness}
\begin{figure}[htbp]
\begin{center}
\includegraphics[width=0.3\linewidth]{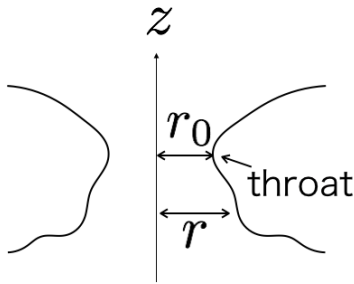}
   \caption{The embedding of the Morris-Thorne type metric. In general, wormholes do not have a mirror symmetry (like the Einstein-Rosen bridge) as long as the flaring out condition is satisfied. }
   \label{fig-morris-thorne-embedding}
\end{center}
\end{figure}

The surface $b=r$ actually describes the throat. The reason for this is obvious from the embedding operation. We can play the same game in Sec \ref{sec-ER-bridge} to get the embedding of the metric \eq{morris-thorne-metric}. Going through the same process in Sec \ref{sec-ER-bridge}, we obtain a differential equation for $z$;
\begin{align}
\frac{\D z}{\D r}=\pm \left(\frac{r}{b(r)}-1 \right)^{-\frac{1}{2}}. \label{morris-thorne-dz/dr}
\end{align}
This differential equation can now be integrated if $b(r)$ is determined. So, $b$ is called the shape function. Obviously, \eq{morris-thorne-dz/dr} diverges when $b=r=:r_0$. Since the schematic picture of \eq{morris-thorne-dz/dr} is similar to \fig{fig-embedding}, one finds that the sphere with radius of $r_0$ describes the throat (\fig{fig-morris-thorne-embedding}).
We denote the throat, $b=r=r_0$, as the minimum surface. As mentioned, at the throat $\D z/\D r=\infty$.
Morris and Thorne further imposed the asymptotically flatness condition which means $\D z/\D r\to 0$ as $r\to\infty$. 

\subsubsection{Flaring-out condition}\label{sec-The flaring-out condition}
In Sec \ref{Embedding wormholes and asymptotically flatness}, we saw \eq{morris-thorne-dz/dr} diverges at the throat. In other words, the inverse function of $z=z(r)$, i.e., $r=r(z)$ satisfies 
\begin{align}
\left. \frac{\D r}{\D z}\right|_{r_0}= \pm \left(\frac{r_0}{b(r_0)}-1 \right)^{\frac{1}{2}}=0.
\end{align}
For a spacetime to be a wormhole, there must be a throat that flares out. 
The flaring-out condition states  
\begin{align}
\frac{\D^2 r}{\D^2 z}=\frac{b-rb'}{2b^2}>0 \label{flare-out-condition}
\end{align}
at and near the throat.

\subsubsection{Absence of the horizon}
For the wormhole to be traversable, there must be no horizons in a spacetime.  
By using the function $\Phi$ in the metric \eq{morris-thorne-metric}, it states 
\begin{align}
\Phi(r) {\rm~ is~ finite~ everywhere}.
\end{align}

\subsubsection{Magnitude of the tension at the throat}
The shape function $b$ gives restrictions on $\rho, \tau$ and $p$ through Eq. (\ref{morris-thorne-p}). A critical restriction is at the throat, $b=r=r_0$. Then $(r-b)\Phi'\to 0$. Reviving $c$ and $G$, this yields the huge tension;
\begin{align}
\tau(r_0)=\frac{c^4}{8\pi Gr_0^2} \sim 5\times 10^{41}\frac{{\rm dyn}}{{\rm cm}^2}\left(\frac{10{\rm m}}{r_0} \right)^2,
\end{align}
if $r_0$ is not too large.

\subsubsection{Exotic matter}\label{sec-Exotic matter}
Besides some of the peculiar features about wormholes noted above, however, the most difficult thing to digest in wormhole physics is the necessity of exotic matters that can violate energy conditions. In general relativity, Morris-Thorne's static spherically symmetric traversable wormholes need stress-energy tensors that violate energy conditions at and near the throat.
To see what happens to the relation between the tension $\tau$ and energy density $\rho$ near the throat, we introduce a dimensionless function $\xi$ as
\begin{align}
\xi:=\frac{\tau-\rho}{|\rho|}=\frac{1}{|b'|}\left(\frac{b}{r}-2(r-b)\Phi'-b'\right). 
\end{align}
Since \eq{flare-out-condition} and $(r-b)\Phi'\to 0$ is satisfied around the throat, $\xi$ reduces to
\begin{align}
\xi|_{r\sim r_0}\simeq \frac{2b^2}{r|b'|}\frac{\D^2 r}{\D^2 z}>0~~ \Leftrightarrow~~ \tau>\rho
\end{align}
at and near the throat. We call the matter which has property $\tau>\rho$ as an {\it exotic matter} because the conventional matter satisfies the null energy condition, $T_{\mu\nu}k^\mu k^\nu\geq0 \Leftrightarrow \tau \leq\rho$.

\subsubsection{Other properties}
Morris and Thorne required some additional conditions on traversable wormholes, tidal forces and a time to pass through wormholes. We do not explain these conditions here since we think conditions that mentioned above (from Sec. \ref{sec-The flaring-out condition} to \ref{sec-Exotic matter}) are the primary problems for wormholes. We refer the reader to Ref. \cite{morris-thorne1988} for the details of tidal forces and time to pass through wormholes.

\subsection{Simple exact solutions and their stability}
If we somehow solved all of the difficulties about wormhole properties discussed above, there is still an important  problem, i.e., the stability of wormhole spacetimes.
Given a spacetime, its stability analysis against gravitational/matter perturbations is a problem with critical importance in the sense that only stable spacetimes may exist in the ``real world''. 

During few decades, after the paper by Einstein and Rosen, several exact solutions to the Einstein equations have been found and they have tunnel structures as described above \cite{Ellis-Bronnikov}. These types of spacetimes are assumed to have a massless scalar field with the opposite sign of its kinetic term to the sign of the Einstein-Hilbert term in the Lagrangian. We often call such a scalar field a {\it ghost} or {\it phantom} scalar field. 
We shall refer to the simplest exact wormhole solution as the Ellis solution (or the Ellis-Bronnikov solution). 

Although wormhole solutions have been known for decades, their stability analysis had not been conducted until quite recently. The first stability analysis was in Ref. \cite{picon}.
In this paper it was showed that the Ellis wormhole is stable against gravitational perturbations in a restricted class which do not change the throat radius.
Subsequently, Shinkai and Hayward showed that the Ellis wormhole is unstable against either a normal and a phantom gaussian pulse of massless scalar field \cite{shinkai-hayward2002}. When a normal (ghost) pulse is injected into the throat, the throat must shrink (inflate).
Gonzalez {\it et al.} have also proved that the Ellis wormhole is unstable against linear and non-linear spherically symmetric perturbations in which the throat is not fixed \cite{gonzalez-guzman-sarbach2009-1, gonzalez-guzman-sarbach2009-2}. They showed that a charged generalization of the wormhole is also unstable \cite{gonzalez-guzman-sarbach-3}.
This unstable feature is invariant in the higher dimensional generalization of the Ellis spacetime \cite{torii-shinkai2013}.

\clearpage

\section{Thin-shell wormholes} \label{Thin-shell wormhole}
As mentioned in section \ref{Wormhole properties in brief}, Morris and Thorne have pioneered qualitative study for static spherically symmetric wormholes.
 
Another class of traversable wormholes has been found by Visser. This class of wormholes can be obtained by a ``cut-and-paste'' procedure~\cite{visser} and such structures are called thin-shell wormholes. Poisson and Visser first
presented stability analysis for spherical perturbations around such thin-shell wormholes, and found that there are stable configurations according to the equation of state of an exotic matter residing on the
throat~\cite{poisson-visser}. The work of Poisson and Visser has been extended in different directions; charged thin-shell wormholes~\cite{Reissner}, thin-shell wormholes constructed by a couple of Schwarzschild spacetimes of different masses \cite{ishak-lake2002} and thin-shell wormhole with a cosmological constant \cite{lobo-crawford2004}. Thin-shell wormholes in cylindrically symmetric spacetimes have also been studied~\cite{cylindricalTSW in Einstein gravity, cylindricalTSW in modified gravity}. Garcia {\it et al.} published a paper about stability for generic static and spherically symmetric thin-shell wormholes~\cite{Garcia+}. Dias and Lemos studied stability in higher dimensional Einstein gravity~\cite{dias-lemos2010}.
Thin-shell wormholes can also be constructed in extended gravity models \cite{Mustafa}.

These wormhole models are easy to construct, which certainly have a wormhole structure. Thin-shell  spacetimes have no 
differentiability for its metric at their throat, i.e., they are only $C^0$, contrary to smooth wormhole models such as the Ellis wormhole, which are smooth ($C^\infty$) at their throat.
Though thin-shell wormholes are not smooth, they indeed have qualitatively common features with smooth wormholes. Hence, investigating the properties of thin-shell wormholes may prompt deeper understanding for the properties of generic wormholes.

In this section, we show the construction and linear stability against perturbations preserving symmetries
of thin-shell wormholes.

\subsection{Junction conditions}
To construct thin-shell wormholes, we first review junction conditions.
Let us consider a hypersurface $\Sigma$ in a spacetime.
$\Sigma$ divides the spacetime into two parts, the one side is $\mathcal M^+$ with coordinates $x^\alpha_+$ and the metric $g^{\alpha\beta}_+$ while the other side is $\mathcal M^-$ with coordinates $x^\alpha_-$ and the metric $g^{\alpha\beta}_-$.
Junction conditions are capable of joining $\mathcal M^+$ and $\mathcal M^-$ at $\Sigma$, so that $g^{\alpha\beta}_\pm$ respect the Einstein equations.

We define the extrinsic curvatures as
\begin{align}
K_{ab}:=n_{\mu;\beta}e^\mu_a e^\beta_b.
\end{align}
We also define $h_{ab}$ as a metric of the line element on $\Sigma$.
Then, for the smooth joint of spacetimes at their boundaries,
\begin{align}
[h_{ab}]=0~~ {\rm and}~~[K_{ab}]=0
\end{align}
are required. 
We have defined the deviation of an arbitrary tensorial function on the both sides of the hypersurface as
$[F]:=(F^+-F^-)|_{\Sigma}$, where $F^+$ and $F^-$ are functions in $\mathcal M^+$ and $\mathcal M^-$, respectively.

On the other hand, if $[K_{ab}]\neq0$, in addition to $[h_{ab}]=0$, there is a stress-energy tensor, $S_{ab}$, which satisfies
\begin{align}
8\pi S_{ab}=- [K_{ab}]+[K] h_{ab}, \label{hypersurface-eom}
\end{align}
The matter with non-zero $S_{ab}$ is confined to the infinitesimally thin surface $\Sigma$, so we call the surface a {\it thin-shell}. Eq. (\ref{hypersurface-eom}) describes the motion for the thin-shell.
The corresponding constraints are given by
\begin{align}
S^{~b}_{a~|b}=- [T_{\alpha\beta}e^\alpha_{a}n^\beta] \label{hamiltonian-cons}
\end{align}
and
\begin{align}
\bar K^{ab}S_{ab}=  [T_{\alpha\beta}n^\alpha n^\beta],
\end{align}
where $\bar K^{ab}:=(K^{ab}_++K^{ab}_-)|_\Sigma/2$.
See Poisson \cite{toolkit} and Barrabes and Bressange \cite{BB1997} for the details of derivation.

\subsection{Construction}
In order to investigate wormhole stability, one must construct thin-shell wormholes. Here, we show how to build them.

First of all, we would like to note that the construction will be
operated in spherically, planarly 
and hyperbolically symmetric spacetimes in $d$-dimensional ($d\geq3$) Einstein gravity with an electromagnetic field and a cosmological constant in bulk spacetimes. Such a rich setup is required for later analysis.

The formalism for maximally symmetric $d$-dimensional thin-shell
wormholes has been developed first by Dias and Lemos
\cite{dias-lemos2010}. We extend their formalism to more general
situations. We obtain wormholes by operating three steps invoking
junction conditions~\cite{toolkit}.

First, consider a couple of $d$ dimensional manifolds, $\mathcal M_\pm$. We assume $d\ge 3$.
The $d$ dimensional Einstein equations are given by
\begin{align}
G_{\mu\nu \pm}+\frac{(d-1)(d-2)}{6}\Lambda_\pm g_{\mu\nu \pm}=8\pi T_{\mu\nu \pm},
\end{align}
where $G_{\mu\nu \pm}$,  $T_{\mu\nu \pm}$ and $\Lambda_\pm$ are Einstein
tensors, stress-energy tensors and cosmological constants in the
manifolds $\mathcal M_{\pm}$, respectively. 
The metrics on $\mathcal M_\pm$ are given by $g^\pm_{\mu\nu}(x^\pm)$. 
The metrics for static and spherically, planar and hyperbolically
symmetric spacetimes 
with $G_{(d-1)(d-2)/2}(d-2,S)$ symmetry
on $\mathcal M_\pm$ are written as 
\begin{align}
&ds_\pm^2=-f_\pm(r_\pm)dt^2_\pm+f_\pm(r_\pm)^{-1}dr^2_\pm+r^2_\pm (d\Omega_{d-2}^k)^2_\pm ,\label{senso}\\
&f_\pm(r_\pm)=k-\frac{\Lambda_\pm r_\pm^2}{3}-\frac{M_\pm}{r_\pm^{d-3}}+\frac{Q_\pm^2}{r_\pm^{2(d-3)}}, \label{static-metric-general}
\end{align}
respectively. 
$M_\pm$ and $Q_\pm$ correspond to the mass and charge parameters in $\mathcal M_\pm$, respectively. $k$ is a constant that determines the 
geometry of the $(d-2)$ dimensional space and takes $\pm 1$ or $0$. 
$k=+1$, $0$ and $-1$ correspond to a sphere, plane 
and a hyperboloid, respectively. $(\D \Omega_{d-2}^k)^2$ is given by
\begin{align}
(\D \Omega_{d-2}^1)^2 =&
\D \theta_1^2 + \sin^2 \theta_1\D\theta_ 2^2 + \ldots + \prod_{i=2}^{d-3}\sin^2 \theta_i \D \theta_{d -2}^2, \nonumber \\
(\D \Omega_{d-2}^0)^2 =&\D \theta_1^2 + \D \theta_2^2 + \ldots + \D \theta_{d - 2}^2, \nonumber\\
(\D \Omega_{d-2}^{-1})^2 =&
\D\theta_1^2 + \sinh^2 \theta_1\D \theta_2^2 + \ldots+ \sinh^2 \theta_1\prod_{i=2}^{d-3}\sin^2\theta_i\D \theta_{d-2}^2.
\end{align}

We should note that by generalized Birkhoff's
theorem~\cite{sato-kodama}, the metric (\ref{senso}) is the unique
solution of Einstein equations of electrovacuum for $k=+1, 0$ and $-1$.

Secondly, we construct a manifold $\mathcal M$ by gluing $\mathcal M_\pm$ at their boundaries. We choose the boundary hypersurfaces 
$\partial \mathcal M_\pm$ as follows:
$\partial \mathcal M_\pm := \{r_\pm =a\ |\ f_\pm(a)>0 \}$,
where $a$ is called a thin-shell radius.
Then, by gluing the two regions $\tilde{\mathcal M}_\pm$ which are defined as
$\tilde{\mathcal M}_\pm := \{r_\pm  \geq a\ |\ f_\pm(a)>0 \}$
with matching their boundaries, $\partial \mathcal M_+=\partial \mathcal M_- := \Sigma$, we can construct a new manifold $\mathcal M$ which has a  wormhole throat at $\Sigma$. $\Sigma$ should be a timelike hypersurface, on which the line element is given by
\begin{align}
\D s_{\Sigma}^2=-d\tau^2+a(\tau)^2(d\Omega_{d-2}^k)^2.
\end{align}
The surface function for $\Sigma$ is given by $\Phi=r-a(\tau)=0$.  
$\tau$ stands for proper time on the junction surface $\Sigma$ whose
position is described by the coordinates
$x^\mu(y^a)=x^\mu(\tau,\theta_1,\theta_2,\ldots,\theta_{d-2})=(t(\tau),a(\tau),\theta_1,\theta_2,\ldots,\theta_{d-2})$,
where Greek indices run over $1,2,\ldots,d$ and Latin indices run over
$1,2,\ldots,d-1$. $\{y^a\}$ 
are 
the intrinsic coordinates on $\Sigma$.

Thirdly, by using the junction conditions, we derive the Einstein equations for the submanifold $\Sigma$. To achieve this, we define unit normals to hypersurfaces $\partial \mathcal M_\pm$. The unit normals are defined by 
\begin{align}
n_{\alpha \pm} := \pm \frac{\Phi_{,\alpha}}{|\Phi^{,\mu}\Phi_{,\mu} |^\frac{1}{2}}. \label{unitnormal}
\end{align}
{\it To construct thin-shell wormholes, we make the unit normals on $\partial
\mathcal M_\pm$ take different signs}, while to construct normal thin
shell models, the unit normals are chosen to be of same signs.
Tangent vectors $e^\alpha_a$ are defined by $\partial x^\alpha/\partial y^a$.
We also define the four-velocity $u_\pm^\alpha$ of the boundary as
\begin{align}
u_\pm^\alpha := e_{\tau \pm}^\alpha=  (\dot{t}_\pm,\dot{a},0,\ldots,0) 
		= \left(\frac{1}{f_\pm(a)}\sqrt{f_\pm(a)+\dot{a}^2}, \dot{a},0,\ldots,0\right),
\end{align}
where $\dot{} := \partial/\partial \tau$ and
$u^{\alpha}u_{\alpha}=-1$ is satisfied. The explicit form of \eq{unitnormal} is
\begin{align}
n_{\alpha \pm}=\pm\left(-\dot{a},\frac{\sqrt{f_\pm+\dot{a}^2}}{f_\pm},0,\ldots,0\right)
\end{align}
and the unit normal satisfies $n_\alpha n^\alpha=1$ and $u^\alpha n_\alpha=0$.

\subsection{Equation of motion for the shell}
The equations for the shell $\Sigma$ are given by \eq{hypersurface-eom}.
The non-zero components of the extrinsic curvature are the following:
\begin{align}
K_\tau^{\tau \pm}=\pm (f_\pm +\dot{a}^2)^{-\frac{1}{2}}(\ddot{a}+\frac{1}{2}f_\pm^\prime), ~~
K_{\theta_1}^{\theta_1 \pm}=K_{\theta_2}^{\theta_2 \pm}=\ldots=K_{\theta_{d-2}}^{\theta_{d-2} \pm}=\pm \frac{1}{a}\sqrt{f_\pm+\dot{a}^2},
\end{align}
where $' := \D/\D a$. 
Since our metrics Eq. (\ref{senso}) are diagonal, $S^i_j$ is also diagonal and written as
\begin{equation}
S^i_j={\rm diag}(-\sigma,p,p,\ldots,p), \label{s,p,p} \\
\end{equation}
where $p$ is the surface pressure (of opposite sign to the surface tension) and $\sigma$ is the surface energy density living on the thin-shell.
Hence, we obtain the explicit form of \eq{hypersurface-eom}:
\begin{align}
& \sigma = -\frac {d - 2}{8\pi a} (A_++ A_ -),  \label{energy density} \\
& p = \frac{1}{8\pi} \left\{ \frac{B_ +}{A_ +} + \frac {B_-} {A_-} + \frac{d - 3}{a}(A_++ A_-)\right\}, \label{surface pressure}
\end{align}
where
\begin{align}
A_\pm(a) := \sqrt{f_\pm+\dot{a}^2} \ ,\ B_\pm(a) := \ddot{a}+\frac{1}{2}f_\pm^\prime.
\end{align}
Note that we deduced a critical property of wormholes that {\it $\sigma$ must be negative}.

The conservation law for the surface stress-energy tensor $S^i_j$ is given by \eq{hamiltonian-cons}.
Since the stress-energy tensor $T^\alpha_\beta$ in the bulk spacetime only contains the electromagnetic field,
$T^\alpha_\beta=Q^2/(8\pi r^{2(d-2)}) {\rm diag} (-1,-1,1,\cdots,1)$,
one can find $T^\alpha_\beta e^{a}_\alpha n^\beta=0$. Hence, \eq{hamiltonian-cons} yields $S^{~b}_{a~|b}=0$ that is
\begin{equation}
\frac{\D }{\D \tau}(\sigma a^{d-2})+p\frac{\D}{\D \tau}(a^{d-2})=0. \label{conservationlaw}
\end{equation}
Eq.~(\ref{conservationlaw}) corresponds to the conservation law. For later convenience for calculations, we recast Eq.~(\ref{conservationlaw}) to 
\begin{align}
\sigma^\prime =-\frac{d-2}{a}(\sigma+p). \label{cons1}
\end{align}
We can get the master equation for the thin-shell throat by recasting Eq.~(\ref{energy density}) as follows:
\begin{align}
\dot {a}^2 + V(a) = 0, \label{energy cons} 
\end{align}
where
\begin{align}
V(a)=-\left(\frac{4\pi a \sigma}{d-2}\right)^2 - \left(\frac{f_+ -f_-}{2}\right)^2 \left(\frac{d-2}{8\pi a \sigma}\right)^2+\frac{1}{2}\left(f_++ f_-\right). \label{general-V(a)}
\end{align}
From Eq.~(\ref{energy cons}), the range of $a$ which satisfies $V(a)\leq 0$ is the movable range for the shell. Since we obtained Eq.~(\ref{energy cons}) by twice squaring Eq.~(\ref{energy density}), there is possibility that we take wrong solutions which satisfy Eq.~(\ref{energy cons}) but do not satisfy Eq.~(\ref{energy density}). 
However, we can show that this is not the case. See Ref. \cite{kh2015} for the proof.
By differentiating Eq.~(\ref{energy cons}) with respect to $\tau$, we get the equation of motion for the shell as
\begin{align}
\ddot a=-\frac{1}{2}V^\prime (a). \label{eom of shell}
\end{align}

Suppose a thin-shell throat be static at $a=a_0$ and its throat radius {\it must} satisfy the relation	
\begin{align}
f(a_0)>0.  \label{f0>0}
\end{align}
This condition is called the horizon-avoidance condition in Ref~\cite{barcelo&visser}. 

We analyze stability against small perturbations preserving symmetries. To determine whether the shell is stable or not against the perturbation, we use Taylor expansion of the potential $V(a)$ around the static radius $a_0$ as
\begin{align}
V(a)=&V(a_0)+V^\prime(a_0)(a-a_0) +\frac{1}{2}V^{\prime \prime}(a_0)(a-a_0)^2+\mathcal O((a-a_0)^3) .\label{V(a)}
\end{align}
From Eqs.~(\ref{energy cons}) and (\ref{eom of shell}), $\dot a_0=0,\ddot a_0=0\Leftrightarrow V(a_0)=0,V^\prime(a_0)=0$ so the potential given by Eq.~(\ref{V(a)}) reduces to
\begin{equation}
V(a)=\frac{1}{2}V^{\prime \prime}(a_0)(a-a_0)^2+\mathcal O((a-a_0)^3). \label{V(a)2}
\end{equation}
The leading term is quadratic as to $a$ and has the co-efficient $V^{\prime \prime}(a_0)$. The sign of the co-efficient makes the form of the potential near static solutions.
Therefore, the stability condition against radial perturbations for the static shell is given by
\begin{equation}
V^{\prime \prime}(a_0)>0. \label{V''>0}
\end{equation}

\subsection{Simplest thin-shell wormhole}
As mentioned before, due to \eq{energy density}, wormholes must have a negative surface energy density, namely, an exotic matter. Hence, the stability of such wormholes depends not only on their geometries but also on exotic matters on $\Sigma$. Here, in the following, we see the simplest thin-shell wormhole made from gluing two Schwarzschild spacetimes and its stability.

The simplest wormhole was first proposed by Visser. As mentioned above, the simplest thin-shell wormhole is constructed from cutting and pasting the couple of identical Schwarzschild spacetimes in the four dimensions, that is, in \eq{static-metric-general} we take $d=4$, $k=1$, $M_\pm=2M$, $Q_\pm=0$ and $\Lambda_\pm=0$ and hence $f(r)_\pm=f(r)=1-2M/r$. Then \eq{general-V(a)} reduces to
\begin{align}
V(a)=f(a)-(2\pi a \sigma)^2, \label{V(a)01}
\end{align}
where $\sigma$ is not specified yet.

\subsection{Stability}
In the simplest setup of thin-shell wormholes, the potential takes the form of \eq{V(a)01}.
In this setup, we study a stability analysis with barotropic fluid and a pure tension matter field.
\subsubsection{Global stability}
First, we learn a global stability analysis with general fluid, i.e., we do not specify an equation of state for the exotic matter residing on the shell. 
We see how the wormhole can be prevented from collapsing to a black hole or expanding infinity.
This case is in the book by Visser \cite{visser}. To begin with, we investigate the asymptotic form of the potential \eq{V(a)01}. The explicit form of \eq{V(a)01} is
\begin{align}
V(a)=1-\frac{2M}{a}-(2\pi a \sigma)^2. \label{potential-general-sch-TSW}
\end{align}
If we have $V(\infty)=1$, the wormhole is prevented from an eternal throat expansion. This condition is explicitly written as 
\begin{align}
V(a\to\infty)=1 \Leftrightarrow \lim_{a\rightarrow 0}(a\sigma)^2=0. \label{sigma-vanish-infinity}
\end{align}
Hence, we can say that $\sigma \rightarrow 0 ~{\rm as}  ~a\to\infty$.
Let us assume Taylor's expansion of $p$ around $\sigma\simeq0$:
\begin{align}
p=p|_{\sigma=0}+\left.\frac{\partial p}{\partial \sigma}\right|_{\sigma=0}\sigma+\mathcal O(\sigma^2)
=:p_{\sigma0}+\beta^2_{\sigma0}\sigma+\mathcal O(\sigma^2). \label{taylor-p}
\end{align}
On the other hand, the conservation law \eq{conservationlaw} is solved for $p$:
\begin{align}
p=-\sigma-\frac{1}{2}\frac{\D \sigma}{\D a}a. \label{conservation-solved-for-p}
\end{align}
From \eq{taylor-p} and \eq{conservation-solved-for-p}, we get
\begin{align}
(\beta^2_{\sigma0}+1)\sigma \simeq-p_{\sigma0}-\frac{1}{2}\frac{\D \sigma}{\D a}a.
\end{align}
For $\beta^2_{\sigma0}\neq-1$, this can be integrated to
$\sigma=Ca^{-2(\beta^2_{\sigma0}+1)}-(\beta^2_{\sigma0}+1)p_{\sigma0}$ 
with a constant $C$. Actually, $p_{\sigma0}$ must vanish because of the asymptotic behavior of $\sigma$. Hence, one obtains
\begin{align}
\sigma=Ca^{-2(\beta^2_{\sigma0}+1)}. \label{sigma-exact}
\end{align}
Due to \eq{sigma-vanish-infinity} and \eq{sigma-exact}, we see the condition that the wormhole is stable against explosion is
\begin{align}
1+2\beta^2_{\sigma0}>0. \label{no-explosion-condition}
\end{align}
We emphasize that this linear equation of state is valid only for large $a$. As a throat moves inwardly from infinity, $\mathcal O(\sigma^2)$ term in \eq{taylor-p} gradually becomes significant.

\begin{figure}[htbp]
\begin{center}
\includegraphics[width=0.45\linewidth]{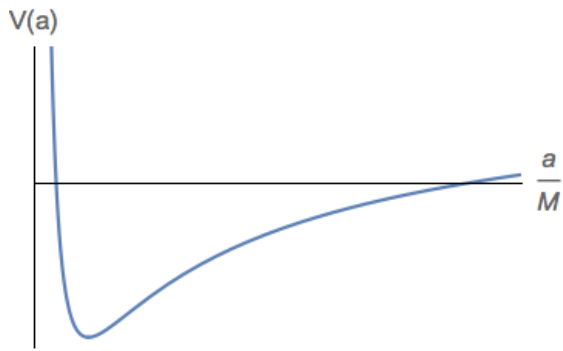}
\caption{Schematic potential of a globally stable wormhole.}
\label{fig-globally-stable}
\end{center}
\end{figure}

Though we have just shown that \eq{no-explosion-condition} is no-explosion condition, we also want to have a no-collapse condition. As mentioned above, we see the contribution of $-2M/a$ is dominant for $a\to 0$. Since there is no hope for $M>0$, we assume $M<0$ for getting no-collapse condition. We also assume that we can use \eq{sigma-exact} also for small $a$.
 This assumption allows us to write the potential  as 
\begin{align}
V(a)= 1+\frac{2|M|}{a}-(2\pi C)^2 a^{-2(2\beta^2_{\sigma0}+1)}.
\end{align}
If
\begin{align}
\beta^2_{\sigma0}<-\frac{1}{4} \label{no-collapse-condition}
\end{align}
is satisfied (and parameters $M$ and $C$ are chosen appropriately), this condition prevents the wormhole throat from collapsing. 
Therefore, the overlap of \eq{no-explosion-condition} and \eq{no-collapse-condition} is fully stable condition:
\begin{align}
-\frac{1}{2}<\beta^2_{\sigma0}<-\frac{1}{4}.
\end{align}
Such a globally stable wormhole is plotted in \fig{fig-globally-stable}.

\subsubsection{Local stability}\label{Barotropic fluid}
Local stability was studied in the paper by Poisson and Visser \cite{poisson-visser}. They assumed barotropic equation of state, $p=p(\sigma)$. Then \eq{cons1} can be integrated as
\begin{align}
\log (a)=-\frac{1}{2}\int \frac{d\sigma}{\sigma+p(\sigma)}. \label{log-a}
\end{align}
Since \eq{log-a} is an equation for $\sigma$, the solution is given by $\sigma=\sigma(a)$. Substituting $\sigma(a)$ into the master equation \eq{V(a)01} reads
\begin{align}
V(a)=f(a)-\left(2\pi a \sigma(a)\right)^2. \label{V(a)-poisson-visser}
\end{align}
Static solutions $a=a_0$ satisfy
\begin{align}
\sigma(a_0):=\sigma_0=-\frac{1}{2\pi a_0}\sqrt{1-\frac{2M}{a_0}} ~~{\rm and}~~
p(a_0):=p_0=\frac{1}{4\pi a_0}\frac{1-M/a_0}{\sqrt{1-2M/a_0}} \label{p001}.
\end{align}
The wormhole is stable if the potential satisfies \eq{V''>0}.
$V''(a_0)$ can be evaluated as
\begin{align}
V^{\prime \prime}(a_0)=-\frac{2}{a^2_0}\left[\frac{2M}{a_0}+\frac{\frac{M^2}{a_0^2}}{1-\frac{2M}{a_0}}+(1+2\beta_0)\left(1-\frac{3M}{a_0}\right)\right],
\end{align} 
where 
\begin{align}
\beta_0 := \left.\frac{\D p}{\D \sigma}\right|_{\sigma_0}.
\end{align}
Hence, the stability condition is
\begin{align}
\beta_0\left(1-\frac{3M}{a_0}\right)<-\frac{1-\frac{3M}{a_0}+\frac{3M^2}{a^2_0}}{2\left(1-\frac{2M}{a_0}\right)}. 
\label{stable-cond-schwarzschild-TSW}
\end{align}
\begin{figure}[htbp]
\begin{center}
\includegraphics[width=0.6\linewidth]{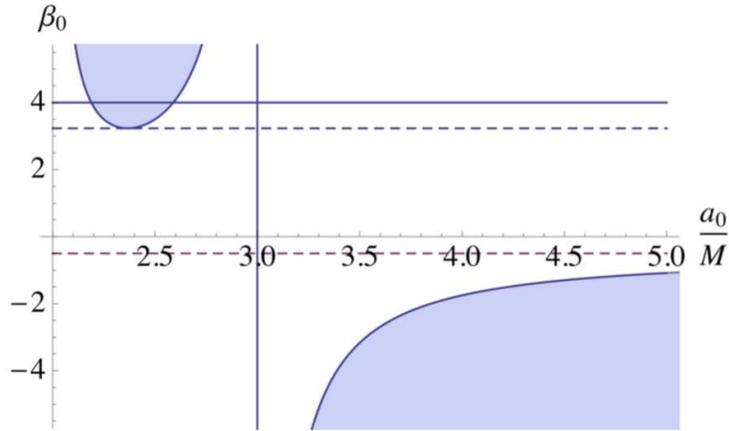}
\caption{If $\beta_0$ is given, one knows the range of $a_0$ for the stable static throat solution. Region $a_0<2M$ is unphysical since such region does not exist in the wormhole. The upper broken line is $\beta_0=(3+\sqrt{3})/2$ while lower is $\beta_0=-1/2$. Shaded regions correspond to stable regions. We draw the horizontal line with $\beta_0=4$ for an example. In this case, wormhole can be stable if $2.19M<a_0<2.59M$.}
\label{fig-stable-region-sch-TSW}
\end{center}
\end{figure}

When $a_0=3M$, it is unstable regardless of the value $\beta_0$ because $V^{\prime \prime}(a_0=3M)=-2/(9M^2)<0$. 
Due to \eq{stable-cond-schwarzschild-TSW}, one can identify the stable range of $a_0$ if $\beta_0$ is given.
Stable regions are depicted as shaded regions in \fig{fig-stable-region-sch-TSW}.
Although there are two values $a_0$ that stand for two extremums of $F(a_0)$, namely, $a_0/M=(3\pm \sqrt{3})/2$, the smaller solution $(3-\sqrt{3})/2$ is an unphysical value because we consider the range $a_0>2M$.

From the above analysis, we can state that if we take a particular value of $\beta_0$, the stability range of $a_0$ is determined. Since $\beta_0$ relates to the equation of state of exotic matters, this statement is equivalent to that the types of exotic matter determine the range of $a_0$ which is the radius of a stable static throat radius.

 We summarize the stability analysis in the assumption of barotropic equation of state:\\
1) There are stable solutions in $\beta_0\geq (3+\sqrt{3})/2$ or $\beta_0<-1/2$.\\
2) No solution in $(3+\sqrt{3})/2>\beta_0>-1/2$ is stable.\\
3) The solution at $a_0=3M$ is unstable regardless of the value of $\beta_0$.\\
In the work of Poisson and Visser, they took the barotropic equation of
state $p=p(\sigma)$ and the stability analysis does not need to specify
the form of the equation of state. As one can see in
\fig{fig-stable-region-sch-TSW}, there is no stable wormhole between
$0<\beta^2_0<1$. $\beta^2_0$ can be interpreted as the square of the
magnitude of the speed of sound for the exotic matter. Poisson and
Visser pointed out that since we do not know  microphysics for exotic
matter, there is no guarantee that $\beta^2_0$ actually is the speed of
sound. Therefore, the region with $\beta^2_0\leq0$ or $\beta^2_0\geq1$ is not a priori ruled out.

\subsubsection{Pure tension}\label{Pure tension}
We will see that what happens if we adopt a pure tension as the exotic matter. The pure tension is an equation of state that is written by
\begin{align}
p=-\sigma ~~~(\sigma:{\rm const.}). \label{pure-tension-eos}
\end{align}
In our wormhole situation, $\sigma$ must be negative-definite because of \eq{energy density}. 
In this setup, due to \eq{p001} and \eq{pure-tension-eos}, we identify the position of the static solution with $a_0=3M$. Then we have $V''(3M)=-4/(27M^2)-4(2\pi \sigma)^2<0$, that is, the wormhole is unstable.

\newpage
\section{Generalized thin-shell wormholes}\label{Generalized Thin-shell wormholes}
In Sec.\ref{Thin-shell wormhole}, we mentioned that there are many generalizations of the thin-shell wormhole introduced by Visser. In this section we review some of its generalizations implemented by several authors.

\subsection{Charged generalization}
\begin{figure}[htbp]
\begin{center}
\includegraphics[width=0.95\linewidth]{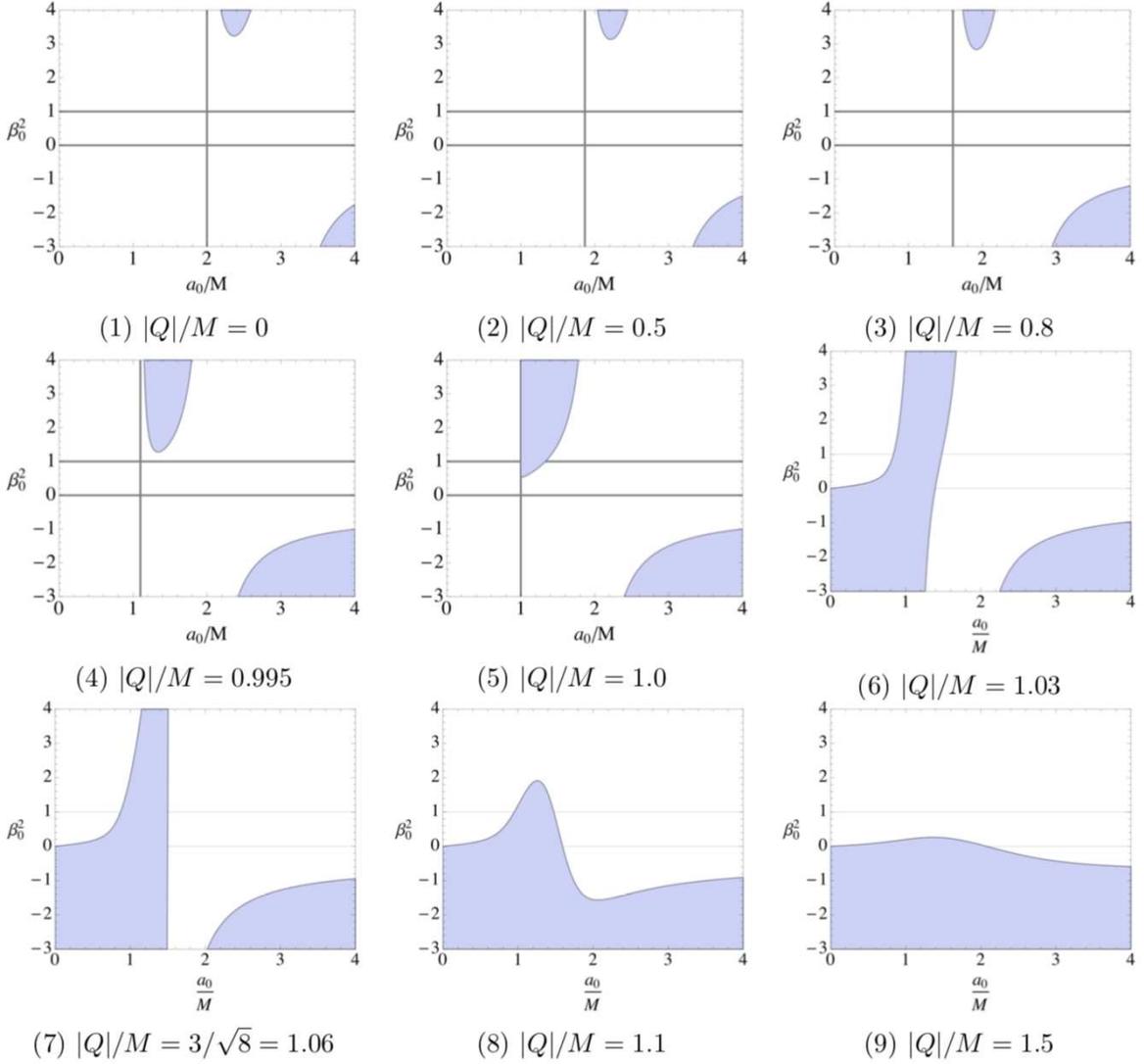}
    \caption{Charged thin-shell wormholes. Shaded regions correspond to
   stable regions. The outer horizon $r_+$ corresponds to
   $r_+/M=1+\sqrt{1-|Q|^2/M^2}$, therefore, the regions inside of the outer horizons (vertical lines) have no physical meaning in these figures when $0\leq |Q|/M\leq 1$. }    \label{fig-ReissnerWH}
\end{center}
\end{figure}

A charged generalization is a theoretically natural extension of the simplest thin-shell wormhole. This is done by Eiroa and Romero \cite{Reissner}. A charged wormhole is constructed from cutting and pasting the couple of identical Reissner-Nordstr\"om spacetimes in  four dimensions.
This situation is reproduced in our formula by putting 
\begin{align}
d=4, k=1, M_\pm=2M, Q_\pm=Q, \Lambda_\pm=0 \nonumber 
\end{align}
into \eq{static-metric-general}.
In this case, stability condition reduces to
\begin{align}
-(1-\frac{3M}{a_0}+\frac{2Q^2}{a_0^2})\beta_0^2>\frac{1}{2(1-\frac{2M}{a_0}+\frac{Q^2}{a_0^2})}\big(1-\frac{3M}{a_0}+\frac{3M^2}{a_0^2}-\frac{Q^2}{a_0M}\big).
\end{align}
Stable regions are presented with each value of charge $|Q|$ in \fig{fig-ReissnerWH}. 
The outer horizon $r_+$ corresponds to 
\begin{align}
\frac{r_+}{M}=1+\sqrt{1-\frac{|Q|^2}{M^2}}.
\end{align}
Therefore, the regions inside of the outer horizons (vertical lines in the figure) have no physical meaning. One finds that stable region appears in $0<\beta^2_0<1$, the sound-speed condition that is supposed to be satisfied for conventional matters.
After the charge reaches the extremal value, $|Q|=M$, there is no longer horizon. So, any static solution $a_0$ can be stable for corresponding $\beta^2_0$.    
We should emphasize that there are {\it always} stable solutions when $1<|Q|/M\leq 3/\sqrt{8}\sim1.06$ {\it regardless of values of} $\beta_0^2$.
%
\subsection{Presence of a cosmological constant}
\begin{figure}[htbp]
\begin{center}
\includegraphics[width=0.85\linewidth]{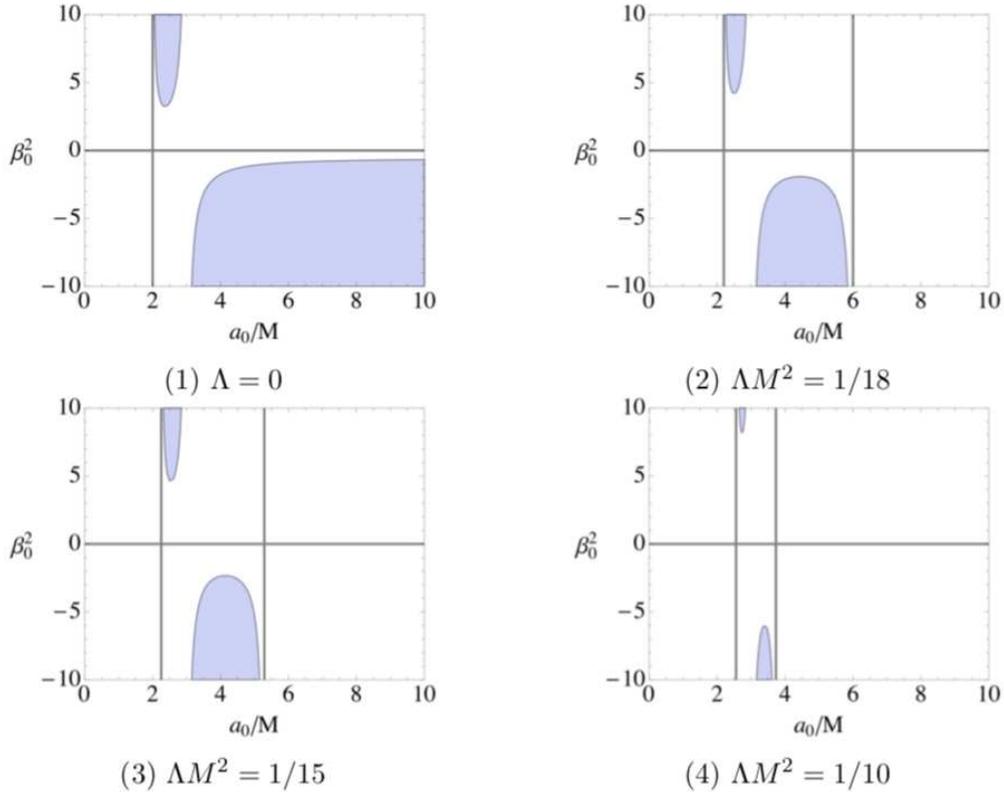}
    \caption{Schwarzschild-de Sitter thin-shell wormholes. The shaded regions indicate stable ones. In figures (2), (3) and (4), the right-side vertical lines correspond to the cosmological horizons. }  
\label{fig-Schwarzschild-de-Sitter-TWS}
\end{center}
\end{figure}

\begin{figure}[htbp]
\begin{center}
\includegraphics[width=0.85\linewidth]{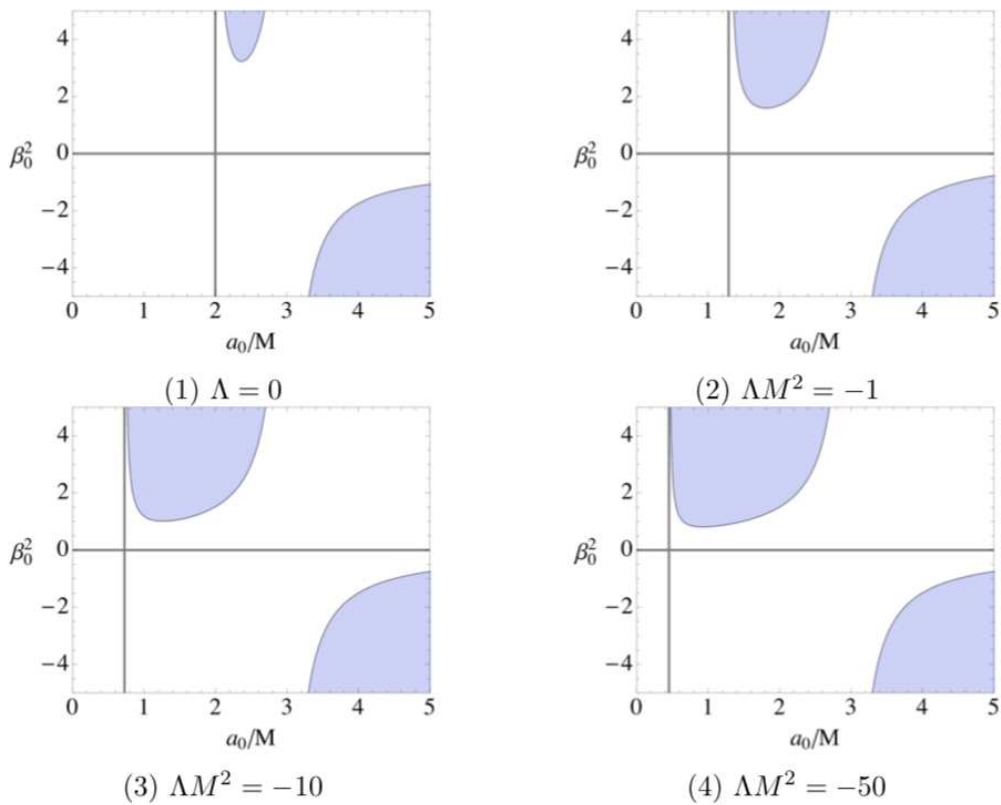}
    \caption{Schwarzschild-anti de Sitter thin-shell wormhole. The shaded regions indicate the stability.}  
\label{fig-Schwarzschild-anti-de-Sitter-TWS}
\end{center}
\end{figure}

Another generalization is implemented in a cosmological sense. Lobo and Crawford developed the construction by introducing a cosmological constant to the simplest wormhole \cite{lobo-crawford2004}. In our notation, such situation is recovered by taking
\begin{align}
d=4, k=1, M_\pm=2M, Q_\pm=0, \Lambda_\pm=\Lambda \nonumber 
\end{align}
in \eq{static-metric-general}.
Stability condition reduces to
\begin{align}
\beta_0^2\left(1-\frac{3M}{a_0}\right)<-\frac{1-3M/a_0+3M^2/a_0^2-\Lambda Ma_0}{2  \left(1-2M/a_0-\Lambda a_0^2/3\right)}. \label{stable-condition-schwarzschild-de-sitterTSW}
\end{align}
Since the ingredient spacetimes are Schwarzschild-(anti) de Sitter, for any $\Lambda$ the event horizon $r_H$ (and also the cosmological horizon for $\Lambda>0$) exists where the equation
\begin{align}
1-\frac{2M}{r}-\frac{\Lambda r^2}{3}=0
\end{align}
holds. Evidently, $\Lambda=0$ corresponds to the Schwarzschild shell wormhole. 
\fig{fig-Schwarzschild-de-Sitter-TWS} and \fig{fig-Schwarzschild-anti-de-Sitter-TWS} show Schwarzschild-de Sitter and Schwarzschild-anti de Sitter thin-shell wormhole, respectively.

\subsubsection{Schwarzschild-de Sitter thin-shell wormhole: $\Lambda>0$}
For each value of $\Lambda M^2$, there are two vertical lines where the denominator diverges in \eq{stable-condition-schwarzschild-de-sitterTSW}. These lines correspond to the event horizon and the cosmological horizon, respectively. Hence, the left (right) side of the event (cosmological) horizon is the unphysical region. For any curves, ``the region above curves lying in $a_0<3M$" and ``the region below curves lying in $a_0>3M$" are stable regions. One finds that  in the case of the Schwarzschild-de Sitter thin-shell wormhole, stability region  decreases with increasing $\Lambda$ relative to the Schwarzschild thin-shell wormhole ($\Lambda=0$).

\subsubsection{Schwarzschild-anti de Sitter thin-shell wormhole:  $\Lambda<0$}
For each value of $\Lambda M^2$, there is a vertical line where the denominator diverges in \eq{stable-condition-schwarzschild-de-sitterTSW}. This vertical line corresponds to the event horizon. Hence, the left side of the event horizon is an unphysical region. Stable regions are similar to those in the case of the Schwarzschild-de Sitter thin-shell wormhole. 
One finds that in the case of the Schwarzschild-anti de Sitter thin-shell wormhole,  the stability region  increases with decreasing $\Lambda$ relative to the Schwarzschild thin-shell wormhole ($\Lambda=0$).
%

\subsubsection{(Anti) de Sitter thin-shell wormhole}
\begin{figure}[htbp]
\begin{center}
\includegraphics[width=0.85\linewidth]{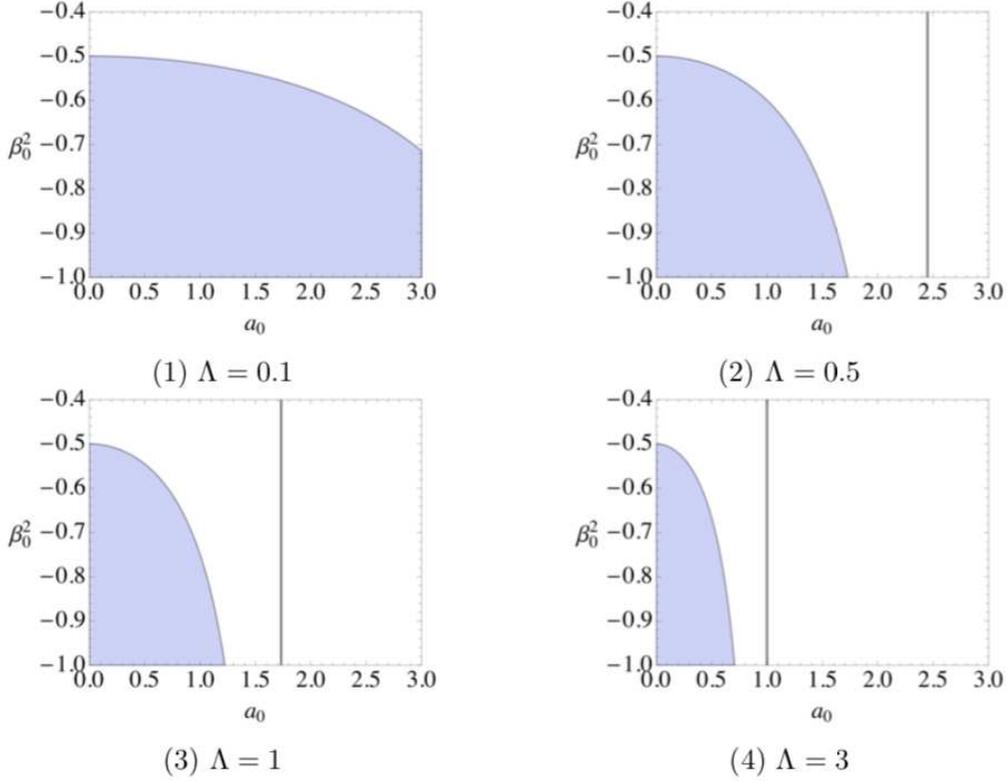}
    \caption{De-Sitter wormholes. A stable region is below the
   curve. Each vertical line represents the cosmological horizon. Hence,
   the right side of the vertical line is an unphysical region.}  
\label{fig-dS-TSW}
\end{center}
\end{figure}

\begin{figure}[htbp]
\begin{center}
\includegraphics[width=0.85\linewidth]{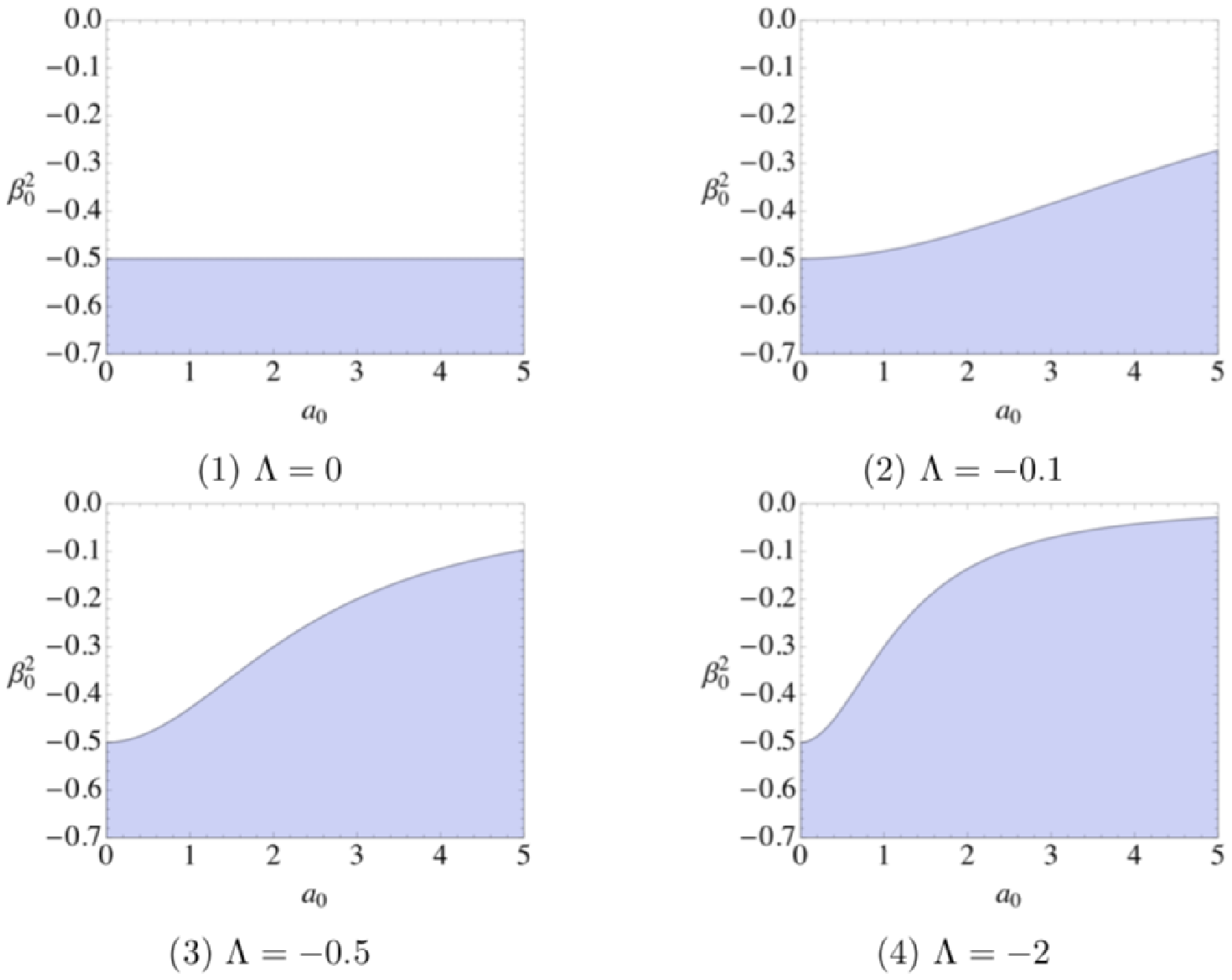}
    \caption{Anti de Sitter wormholes. Stable region is below the curve. (1) describes a wormhole in the Minkowski spacetime.}  
\label{fig-AdS-TSW}
\end{center}
\end{figure}

We can also construct (anti) de Sitter wormhole just by putting $M\to0$ in \eq{stable-condition-schwarzschild-de-sitterTSW}:
\begin{align}
\beta_0^2<-\frac{1}{2 (1-\Lambda a_0^2/3)}. 
\end{align}
See \fig{fig-dS-TSW} and \fig{fig-AdS-TSW} for stable equilibrium.
%
%

\subsection{Non-$Z_2$ symmetric case}
\begin{figure}[htbp]
\begin{center}
\includegraphics[width=0.85\linewidth]{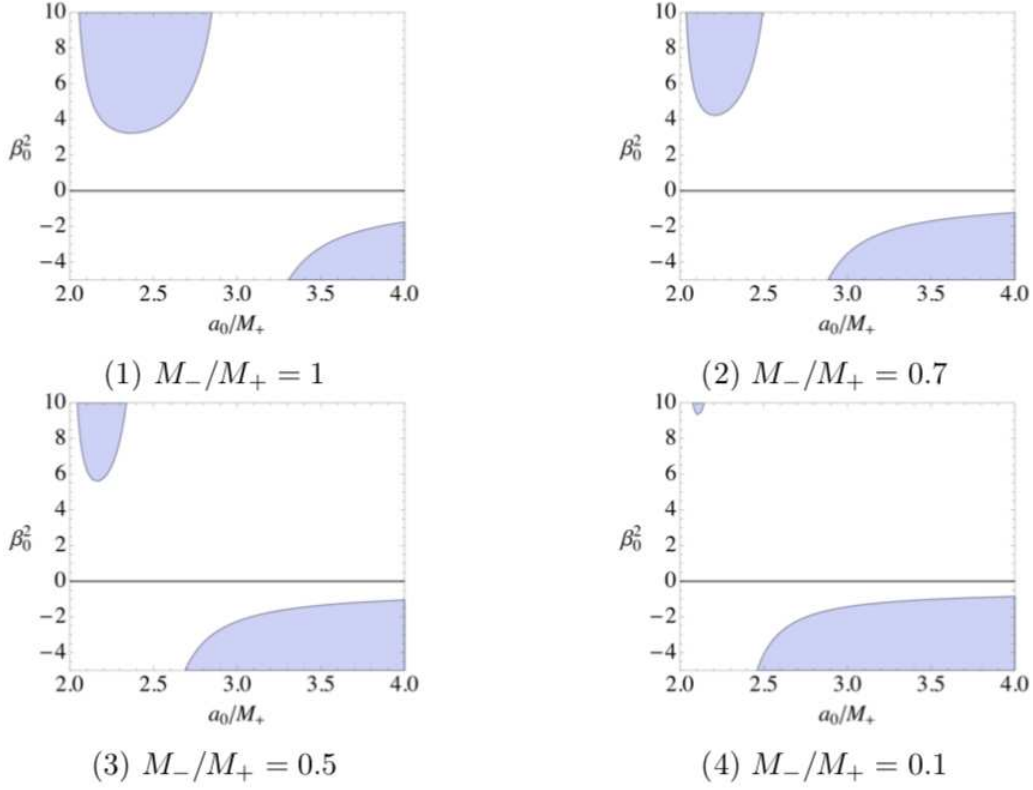}
    \caption{Non-$Z_2$ symmetry. The shaded regions represent stability. The area of stable regions decreases with increasing the difference between $M_+$ and $M_-$. The horizontal axis is normalized by $M_+$.}  
\label{fig-non-Z2-TSW}
\end{center}
\end{figure}

The simplest thin-shell wormhole is made by gluing a couple of Schwarzschild spacetimes with $M_+=M_-=:M$. Ishak and Lake expanded this analysis into the different-mass case, i.e., $M$ is not continuous at $\Sigma$. See \cite{ishak-lake2002} for detail. This situation corresponds to
\begin{align}
d=4, k=1, Q_\pm=\Lambda_\pm=0
\end{align}
in \eq{static-metric-general}. In four dimensions, if we transform $M_\pm$ as $M_\pm\to 2M_\pm$ in \eq{static-metric-general}, then, $M_\pm$ is interpreted as the ADM mass.
In this situation, $V^{\prime \prime}(a_0)$ reduce to the following form:
\begin{align}
V^{\prime \prime}(a_0)=-\mathcal A
-\frac{2}{(d-2)^2}\left(\mathcal C^2 + \mathcal B \mathcal D \right) 
-\frac{(d-2)^2}{2}\left(\mathcal F^2+\mathcal E \mathcal G \right),  \label{d-dim-non-z2-potential}
\end{align}
where
\begin{align}
\mathcal A:=& (d-2)(d-3)\frac{\bar M}{a_0^{d-1}} , ~~
\mathcal B:= -\frac{d-2}{2}\delta_1, ~~
\mathcal C:= -\frac{d-2}{2}\frac{\delta_2}{a_0}, \\
\mathcal D:=& \frac{(d-2)(d-3)}{2a_0^2}(\delta_2-\delta_1) + \frac{(d-2)^2}{2a_0^2}(\delta_2-\delta_1)\beta_0^2, ~~
\mathcal E:= -\frac{2}{(d-2)}\frac{m(a_0)}{a_0\delta_1},\\
\mathcal F:=& -\frac{2}{(d-2)}\frac{m^\prime(a_0)}{a_0\delta_1} + \frac{2}{(d-2)}\frac{m(a_0)}{a_0^2}\frac{\delta_1+\delta_2}{\delta_1^2},\\
\mathcal G:=&  -\frac{2}{(d-2)}\frac{m^{\prime \prime}(a_0)}{a_0\delta_1} +\frac{4}{d-2}\frac{m^\prime(a_0)}{(a_0\delta_1)^2}(\delta_1+\delta_2) \nonumber \\
+& 2\frac{d-3}{d-2}\frac{m(a_0)}{a_0^3\delta_1} -2\frac{d-5}{d-2}\frac{m(a_0) \delta_2}{a_0^3\delta_1^2} -\frac{2m(a_0)}{(a_0\delta_1)^2} \frac{1}{a_0}(\delta_2-\delta_1)  \beta_0^2 
-\frac{4}{(d-2)}\frac{m(a_0)}{(a_0\delta_1)^3}(\delta_1+\delta_2)^2. \label{d-dim-non-z2-potential-G}
\end{align}
We have defined $m, \bar x$ and $\hat x$ as
\begin{align}
m(a):= \frac{\hat M}{a^{d-4}}, ~~
\bar x := \frac{x_++x_-}{2},~~ \hat x := \frac{x_+-x_-}{2}.
\end{align}
$\delta_{1,2}$ was also defined through the equations:
\begin{align}
\sigma(a_0) &= -\frac{d-2}{8\pi a_0}(A_{+}(a_0)+A_{-}(a_0)) =: -\frac{d-2}{8\pi a_0} \delta_1, \\
p(a_0) &=\frac{1}{8\pi a_0}\left\{ \left(\frac{B_{+}(a_0)}{A_{+}(a_0)}+\frac{B_{-}(a_0)}{A_{-}(a_0)}\right)a_0 +(d-3)\{A_{+}(a_0)+A_{-}(a_0)\}  \right\} \nonumber \\
&=: \frac{1}{8\pi a_0}(\delta_2+(d-3)\delta_1).
\end{align}
Putting  $d=4$ into \eq{d-dim-non-z2-potential} recovers the condition $V^{\prime \prime}(a_0)>0$ by Ishak and Lake. Stable equilibrium is depicted in \fig{fig-non-Z2-TSW}. We found that the area of stable regions decreases with increasing the difference between $M_+$ and $M_-$.
%

\subsection{In higher dimensions}
\begin{figure}[htbp]
\begin{center}
\includegraphics[width=0.85\linewidth]{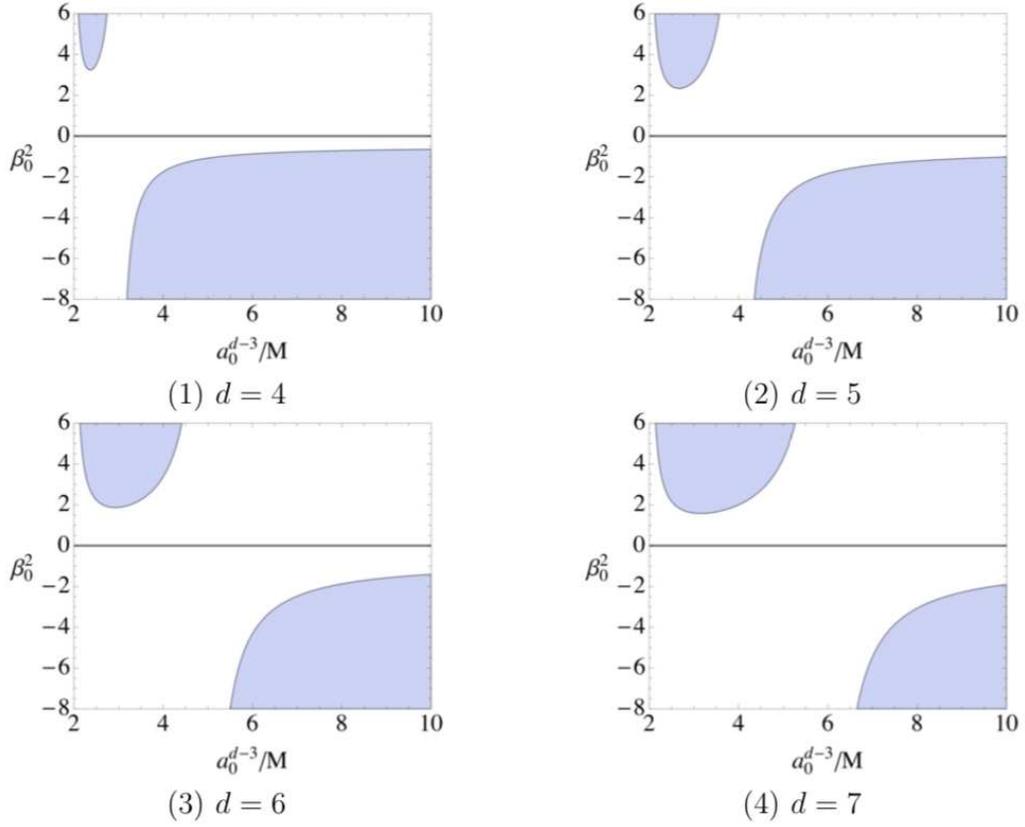}
    \caption{A higher dimensional generalization. The shaded regions represent the stability.}  
\label{fig-high-dim-sch-TSW}
\end{center}
\end{figure}

A higher dimensional generalization is introduced by Dias and Lemos in \cite{dias-lemos2010}. We show a spherically symmetric shell wormhole in arbitrary dimensions, say, a Tangherlini shell wormhole. 
In \eq{static-metric-general}, this situation corresponds to
\begin{align}
k=1, M_\pm= 2M, Q_\pm=\Lambda_\pm=0.
\end{align}
Putting these values into \eq{d-dim-non-z2-potential} - \eq{d-dim-non-z2-potential-G}, we have the stable condition. \fig{fig-high-dim-sch-TSW} shows the stable equilibrium. We find that the area of the stable regions increases with $d$.
%

\clearpage
\section{Pure tension wormholes in Einstein gravity}\label{Pure tension wormholes in Einstein gravity}

\subsection{Einstein gravity} \label{Higher dimensional gravitational theories}
Though general relativity is the most successful simple gravitational theory, many alternative theories have also been theoretically investigated. Particularly, gravity models in four or higher dimensions have been investigated in many aspects:

The notion of higher dimensional spacetimes was first introduced by Kaluza and Klein \cite{KK}. They found that the gravitational field and the electromagnetic field can be unified in five dimensional
spacetimes. In their work, the length of the fifth dimension is confined to a very small scale.

The gauge/gravity correspondence conjectured by Maldacena \cite{gauge/gravity duality} generally 
sheds light on structures in the anti de Sitter spacetime in higher dimensions. 

In the end of the twentieth century, Randall and Sundrum proposed an
idea that we perhaps live in a (3+1) dimensional brane in the (4+1)
dimensional spacetime whose extra dimension spreads widely with a
negative cosmological constant, which is called a brane-world model
\cite{randall-sundrum1}. In this model, the bulk five dimensional
spacetime is the anti de Sitter spacetime. 

In a broader context, candidate theories for quantum gravity, such as superstring theory and M-theory, entail higher dimensional spacetimes.

The most natural and the simplest generalization into higher dimension is the higher dimensional generalization of general relativity. This generalization is taken by writing the gravitational action in vacuum as
\begin{equation}
S_{{\rm E}}=\frac{1}{2\kappa_d^2}\int \D^dx\sqrt{-g}R ~~~(d\geq3),
\end{equation}
where $\kappa_d:=\sqrt{8\pi G_d}$ and $G_d$ is a $d$-dimensional gravitational constant. 
The Ricci scalar $R$ is associated with the $d$-dimensional metric. Variation of the action with respect to the metric gives the same form of general relativity, $G_{\mu\nu}=0$.

\subsection{Advantage of use of pure tension}
Pure tension branes, whether the tension is positive or negative, have particular interest because they have Lorentz invariance and have no intrinsic dynamical degrees of freedom. In the context of stability, 
pure negative-tension branes have no intrinsic instability by their own, although they violate the weak energy condition. This is in contrast with the Ellis wormholes, for which a phantom scalar
field is assumed as a matter content and it suffers the so-called ghost instability because of the kinetic term of a wrong sign~\cite{shinkai-hayward2002, gonzalez-guzman-sarbach2009-1, gonzalez-guzman-sarbach2009-2}.

The construction of traversable wormholes by using negative tension
branes have first been proposed by Barcel\'o and Visser
\cite{barcelo&visser}. 
We proved that a wormhole with a pure tension brane analyzed in Sec. \ref{Pure tension} was unstable. Barcel\'o and Visser investigated pure tension wormholes in more general construction: 
They analyzed the dynamics of spherically symmetric traversable wormholes obtained by operating the cut-and-paste procedure for negative tension 2-branes (three dimensional timelike singular 
hypersurface) in four dimensional spacetimes. 
They found stable brane wormholes constructed by pasting a couple of
Reissner-Nordstr\"om-(anti) de Sitter spacetimes. In their work
\cite{barcelo&visser}, the both of the charge and negative cosmological constant are important to sustain such stable wormholes. And in most cases, a negative cosmological constant tends to
make the black hole horizons smaller. However, in exceptional cases, one
can obtain wormholes with a vanishing cosmological constant, if 
the absolute value of the charge satisfies a certain condition.\\

\subsection{Pure tension wormholes in Einstein gravity}
Here,  
we concentrate on stability of thin-shell wormholes against radial perturbations.
The radial perturbation is important in the context of stability analysis in the following reasons:
(i) Since the stability analysis against radial perturbations is
much simpler than nonradial perturbations which entail
gravitational radiation, it is a natural first step to investigate 
radial stability of wormhole models.
For thin-shell models, the stability analysis against radial perturbations is particularly simple.
(ii) The previous study suggests that the radial
perturbation of wormhole spacetimes is most dangerous: The
paper of Ref. \cite{picon} showed that the Ellis wormhole is
stable against metric perturbations including nonradial perturbations
which do not change the throat radius.
Subsequently, the Ellis wormhole turned out to be unstable 
against radial perturbation which changes the radius of the throat. 
The throat must shrink or inflate \cite{shinkai-hayward2002}.
From the above, we might say that for the Ellis wormhole, the radial
perturbation which changes the radius of the throat 
is the most ``dangerous'' perturbation, as mentioned in
Bronnikov's book \cite{bronnikov-rubin}. One can expect that this
property applies to not only the Ellis wormhole but also 
other wormhole solutions.

The effects caused by the electromagnetic field on the stability of a
thin-shell wormhole are not well known. One may wonder whether the existence of the
electric charge of wormholes can stabilize the wormholes. 
Therefore, it is worth studying electrically charged thin-shell wormholes.

In recent several years, the possibility of stable wormholes 
in various theories of modified gravity
 has gathered attention and has been extensively investigated by many
 authors \cite{cylindricalTSW in modified gravity, TSWs in modified, kkk2011}. However, the possibility of wormholes whose exotic matter
 is a pure tension brane has not yet fully been checked in those
 theories. To study such possibility, it is important and necessary to
fully understand the existence and stability of all kinds of $Z_2$ 
symmetric 
Reissner-Nordstr\"{o}m-(anti) de Sitter
thin-shell wormholes in higher dimensional pure Einstein-Maxwell theory.
 
In this section, we investigate negative-tension branes as thin-shell
wormholes in spherical, planar 
and hyperbolic
symmetries in $d$ dimensional Einstein gravity with an
electromagnetic field and a cosmological constant in bulk
spacetimes. In spherical geometry, we find the higher dimensional counterpart of Barcel\'o and Visser's wormholes which are stable against spherically symmetric perturbations. As the number of dimensions increases, larger charge is allowed to construct such stable wormholes without a cosmological constant. Not only in spherical geometry, but also in planar and hyperbolic geometries, we find static wormholes which are stable against perturbations preserving symmetries. 

\subsection{Effective potential}
From now on, for simplicity, we assume $Z_2$ symmetry, that is, we assume
$M_+=M_-$, $Q_+=Q_-$ and $\Lambda_+=\Lambda_-$ and, hence, $f_+(r)=f_-(r)$.
We denote $M_+=M_-=M$, $Q_+=Q_-=Q$, $\Lambda_+=\Lambda_-=\Lambda$ and 
$f(r):=f_+(r)=f_-(r)$. 
We investigate wormholes which consist of a negative-tension brane. From Eqs.~(\ref{energy density}) and (\ref{surface pressure}), the surface energy density and surface pressure for the negative-tension brane are represented as
\begin{align}
& \sigma = -\frac{d-2}{4\pi a}A=\alpha,  \label{energy density01} \\
& p = \frac{1}{4\pi} \left( \frac{B}{A} + \frac{d - 3}{a}A\right)=-\alpha, \label{surface pressure01}
\end{align}
where $\alpha$ is a constant with a negative value. The
effective potential reduces to
\begin{align}
V(a)=f(a)-\left(\frac{4\pi \alpha}{d-2}\right)^2a^2. \label{V(a)02}
\end{align}

\subsection{Static solutions and stability criterion}
The present system may have static solutions $a=a_{0}$. 
The equation for the static solutions is given by
\begin{align}
2ka_0^{2(d-3)} - (d-1)Ma_0^{d-3} + 2(d-2)Q^2=0. \label{static sol02}
\end{align}
The stability condition for wormholes was shown before as $V^{\prime \prime}(a_0)>0$. The corresponding stability condition is
\begin{align}
V^{\prime \prime}(a_0)>0~ \Leftrightarrow~ (d-3)\left(4k-(d-1)\frac{M}{a_0^{d-3}}\right)<0. \label{stablecondition01}
\end{align}
As one can see, since the static solutions Eq.~(\ref{static sol02}) and stability condition Eq. (\ref{stablecondition01}) do not contain $\Lambda$, the cosmological constant only affects the horizon-avoidance condition of Eq.~(\ref{f0>0}). By studying both the existence of static solutions and stability condition, we can search static and stable wormholes.

For $d\geq4, k\neq 0$ and $M\neq0$,
Eq.~(\ref{static sol02}) is a quadratic equation. The static solutions are then given by 
\begin{align}
a_{0\pm}^{d-3}= \frac{d-1}{4k}M( 1\pm b ), \label{k=+-1:static sol}
\end{align}
where
\begin{align}
b:= \sqrt{1-k\frac{q^2}{q_c^2}}, ~ q:=\frac{|Q|}{|M|}, ~q_c:=\frac{(d-1)}{4\sqrt{d-2}}. \label{definition}
\end{align}
Combining Eqs.~(\ref{stablecondition01}) and (\ref{k=+-1:static sol}),
we can see that for $b=0$, the positive and negative sign solutions 
coincide and their stability depends on higher order terms.

As a specific example of stability analysis, we review for $d\geq4, k=1$ and $M\neq0$ case below.
For $b>0$ and $k=+1$, the negative sign solution is stable, while the positive sign solution is unstable. 
When $0<b<1\ \Leftrightarrow  0 < q<q_c$, there are two static solutions
determined by Eq. (\ref{k=+-1:static sol}).
The stability condition is satisfied if we take the negative sign solution of Eq. (\ref{k=+-1:static sol}). The positive sign solution is unstable. Since the radius of the static wormhole must be positive, we must have $M>0$.
In this stable case, one can achieve this situation if and only if
\begin{align}
\frac{1}{2}<q<q_c \label{ABCD}
\end{align}
is satisfied. 
Therefore, we can construct a stable thin-shell wormhole without $\Lambda$ when the condition Eq.~(\ref{ABCD}) is satisfied. So, the extremal or subextremal charge $q \leq1/2$ of the Reissner-Nordstr\"om spacetime cannot satisfy Eq.~(\ref{ABCD}). We can reconfirm the previous result by taking $d=4$ and $M=2m$ for Eq.~(\ref{ABCD}) as
\begin{align}
1<\left(\frac{|Q|}{m}\right)^2 < \frac{9}{8}.
\end{align}
This coincides with the previous result by Barcel\'o and
Visser~\cite{barcelo&visser}. From Eq.~(\ref{ABCD}), as $d$ increases,
larger charge is allowed to construct a stable wormhole without
$\Lambda$. This class of wormholes are constructed by pasting a couple
of over-charged higher dimensional Reissner-Nordstr\"om spacetimes.

\begin{figure}[ht]
  \begin{center}
    \begin{tabular}{c}

      \begin{minipage}{0.5\hsize}
        \begin{center}
          \includegraphics[scale=0.45]{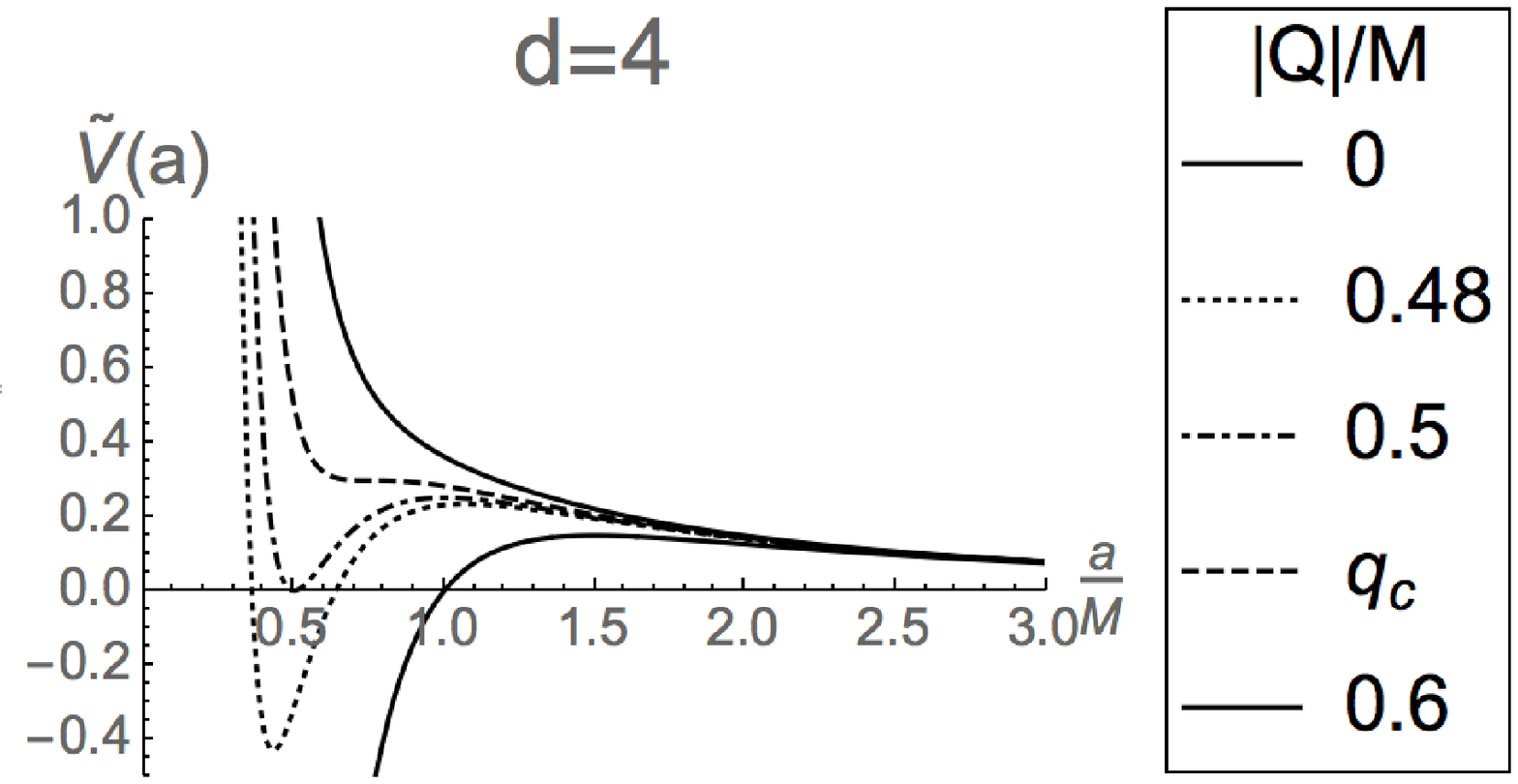}
          \hspace{1.6cm} (1) 
        \end{center}
      \end{minipage}

      \begin{minipage}{0.5\hsize}
        \begin{center}
          \includegraphics[scale=0.45]{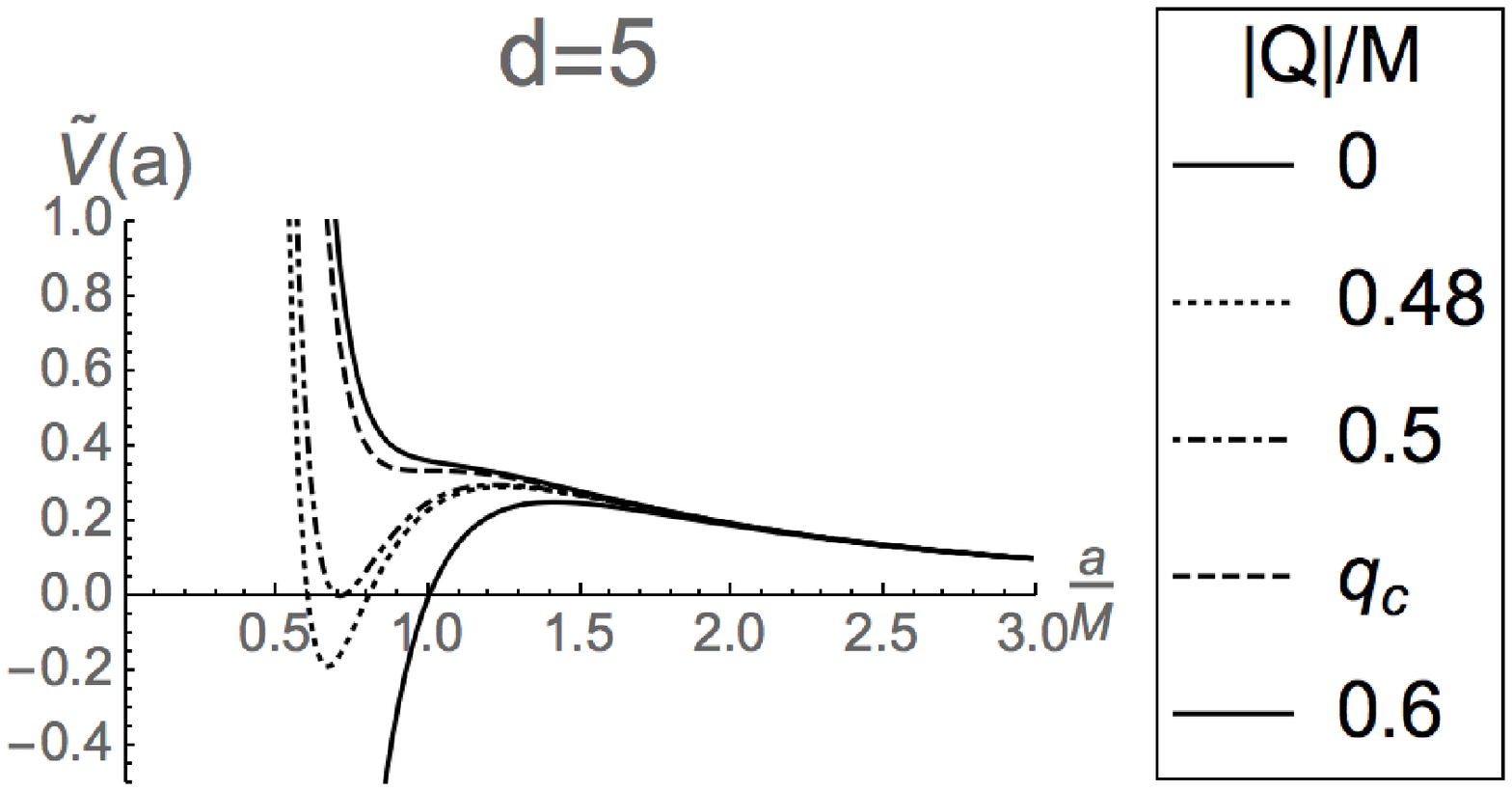}
          \hspace{1.6cm} (2) 
        \end{center}
      \end{minipage}

    \end{tabular}
    \caption{(1) The potential $\tilde V (a)$ for $d=4$, $k=+1$ and $M>0$. The dashed line is the potential for the critical value defined in Eq.~(\ref{definition}). (2) The potential for $d=5$, $k=+1$ and $M>0$.}  
\label{fig-potentials}
  \end{center}
\end{figure}

The following transformation helps us to understand the potentials:
\begin{align}
\tilde V (a):=
\frac{k}{a^2}-\frac{M}{a^{d-1}}+\frac{Q^2}{a^{2(d-2)}} 
\end{align}
so that 
\begin{align}
\dot a^2+V(a)=0 \Leftrightarrow \left( \frac{d\ln a}{d\tau} \right)^2+\tilde V(a)
=\Lambda_{\rm eff},
\end{align}
where $\Lambda_{\rm eff}$ is a constant defined by
\begin{align}
\Lambda_{\rm eff}:=\frac{\Lambda}{3}+\left(\frac{4\pi\alpha}{d-2}\right).
\end{align}
The potentials $\tilde{V}(a)$ are plotted in Figs.~\ref{fig-potentials} for $d=4$ and $d=5$, respectively. 

In the above, we discussed the stability analysis for $k=1$. The analysis for $k=0$ and $k=-1$ has also been completed. See Ref. \cite{kh2015} for the detail of the analysis. 
We summarize the stability analysis of the wormholes for all combinations of $k, M$ and $q$ 
in tables \ref{table d=3} -- \ref{table k=0}. See Ref. \cite{kh2015} for the definitions of $\lambda, R(d), H_\pm(d,q)$, $I(d,q), N(d,q), S(d,q)$ and $J(d,q)$ are given in Ref. \cite{kh2015}.

\clearpage
%
\begin{table*}[htbp]
\begin{center}
\caption{The existence and stability of static wormholes in three dimensions.}
\label{table d=3}
\begin{tabular}{|c|c|c|c|}\hline
 & Static solution & Horizon avoidance & Stability \\ \hline 
$k-M+Q^2=0$ & $\forall a_0>0$ & Satisfied & Marginally stable \\ \hline
$k-M+Q^2 \neq0$ & None & -- & -- \\ \hline
\end{tabular}
\end{center} 
\end{table*}

\vspace{-30pt}

\begin{table*}[htbp]
\begin{center}
\caption{The existence and stability of static wormholes in spherical symmetry in four and higher dimensions. 
\label{table k=+1} }
\begin{tabular}{|c|c|c|c|c|}\hline
\multicolumn{2}{|c|}{} & {Static solution} & Horizon avoidance & Stability \\ \hline\hline
& $q=0$ & $[(d-1)M/2]^{1/(d-3)}$ & $\lambda<H_{+}(d,0)$ & Unstable \\ \cline{2-5}
$M>0$ & $0<q<q_c$ & $a_{0\pm}$ &  $\lambda<H_{\pm}(d,q)$ & \shortstack{$a_{0-}$: Stable\\ $a_{0+}$: Unstable} \\
\cline{2-5} & $q=q_c$ & $[(d-1)M/4]^{1/(d-3)}$  & $\lambda<R(d)$ & Unstable \\ \cline{2-5}
& $q_{c}<q$ & None & -- & --  \\ \hline
\multicolumn{2}{|c|}{$M\leq0$}  & None & -- & --   \\ \hline
\end{tabular}
\end{center}
\end{table*}

\vspace{-30pt}

\begin{table*}[htbp]
\begin{center}
\caption{The existence and stability of static wormholes in planar 
symmetry in four and higher dimensions.}
\label{table k=0}
\begin{tabular}{|c|c|c|c|c|}\hline
\multicolumn{2}{|c|}{}&{Static solution} & Horizon avoidance& Stability 
 \\ \hline\hline 
$M>0$& $q=0$ & None & -- & -- \\ \cline{2-5}
& $q>0$ &$[2(d-2)q^{2}M/(d-1)]^{1/(d-3)}$ & $\lambda<J(d,q)$ & Stable \\ \hline
$M<0$  & $q\geq 0$ & None & -- & --\\ \hline 
$M=0$ & $Q=0$ &$\forall a_0>0$ & Satisfied & \shortstack{Marginally \\stable} \\ \cline{2-5}
& $|Q|>0$ & None & -- & --\\ \hline
\end{tabular}
\end{center} 
\end{table*}

\begin{table*}[htbp]
\begin{center}
\caption{The existence and stability of static wormholes in hyperbolic symmetry in four and higher dimensions.}
\label{table k=-1}
\begin{tabular}{|c|c|c|c|c|}\hline
 \multicolumn{2}{|c|}{}&{Static solution} & Horizon avoidance& Stability 
 \\ \hline\hline 
$M>0$ & $q=0$ & None & -- & -- \\ \cline{2-5}
& $q>0$ & $a_{0-}$ & $\lambda<I(d,q)$ & Stable \\ \hline
$M<0$  & $q=0$ & $[(d-1)|M|/2]^{1/(d-3)}$  & $\lambda <N(d,0)$  & Stable\\ \cline{2-5}
&$q>0$ & $a_{0+}$ & $\lambda<N(d,q)$& Stable \\ \hline
$M=0$  & $Q=0$ & None & -- & -- \\ \cline{2-5}
& $|Q|>0$ & $[\sqrt{d-2}|Q|]^{1/(d-3)}$ & $\Lambda/3<S(d,q)$ & Stable \\ \hline
\end{tabular}
\end{center} 
\end{table*}

\clearpage

\section{Pure tension wormholes in Einstein-Gauss-Bonnet gravity}\label{Pure tension wormholes in Einstein-Gauss-Bonnet gravity}

\subsection{Einstein-Gauss-Bonnet gravity} \label{Einstein-Gauss-Bonnet gravity} 
The action of the $d(\ge 5)$-dimensional Einstein-Gauss-Bonnet gravity in vacuum is given by 
\begin{equation}
S=\frac{1}{2\kappa_d^2}\int \D^dx\sqrt{-g}\biggl(R-2\Lambda+\alpha{L}_{\rm GB}\biggl), \label{action}
\end{equation}
where $\Lambda$ is the cosmological constant. 
The Gauss-Bonnet term ${L}_{\rm GB}$ is defined by the following special combination of the Ricci scalar $R$, the Ricci tensor $R_{\mu\nu}$ and the Riemann tensor $R^\mu{}_{\nu\rho\sigma}$:
\begin{equation}
{L}_{\rm GB}:=R^2-4R_{\mu\nu}R^{\mu\nu}+R_{\mu\nu\rho\sigma}R^{\mu\nu\rho\sigma}.
\end{equation}
The Gauss-Bonnet term appears in the action as the ghost-free quadratic curvature correction term in the low-energy limit of heterotic superstring theory in ten dimensions (together with a dilaton)~\cite{Gross}.
In this context, the coupling constant $\alpha$ is regarded as the inverse string tension and positive definite. 

For this reason, we assume $\alpha > 0$ throughout this article.
In addition, we put another conservative assumption $1+4{\tilde\alpha}{\tilde\Lambda}>0$, where
\begin{equation}
{\tilde\Lambda}:=\frac{2\Lambda}{(d-1)(d-2)},\qquad {\tilde\alpha}:=(d-3)(d-4)\alpha,
\end{equation}
so that the theory admits maximally symmetric vacua, namely Minkowski,
de~Sitter (dS), or anti de Sitter (AdS) vacuum solutions.
Although there exists a maximally symmetric vacuum for $1+4{\tilde\alpha}{\tilde\Lambda}=0$, we do not consider this case for simplicity.

The variation of the action (\ref{action}) with respect to the metric $g^{\mu\nu}$ gives the following vacuum Einstein-Gauss-Bonnet equations:
\begin{equation}
{G}^\mu{}_\nu +\alpha {H}^\mu{}_\nu +\Lambda \delta^\mu{}_\nu=0, \label{beq}
\end{equation}
where
\begin{align}
{G}_{\mu\nu}:=&R_{\mu\nu}-\frac{1}{2}g_{\mu\nu}R, \\
{H}_{\mu\nu}:=&2\Bigl(RR_{\mu\nu}-2R_{\mu\alpha} R^\alpha{}_\nu -2R^{\alpha\beta} R_{\mu\alpha\nu\beta} + R_\mu{}^{\alpha\beta\gamma}R_{\nu\alpha\beta\gamma} \Bigr) - \frac{1}{2} g_{\mu\nu} {L}_{\rm GB}.
\end{align}
The tensor ${H}_{\mu\nu}$ obtained from the Gauss-Bonnet term does not give any higher-derivative term and ${H}_{\mu\nu}\equiv 0$ holds for $d\le 4$.
As a result, Einstein-Gauss-Bonnet gravity is a second-order quasi-linear theory as Einstein gravity is.

\subsection{Pure tension wormholes in Einstein-Gauss-Bonnet gravity}
In 2011, Kanti, Kleihaus, and Kunz numerically constructed four-dimensional spherically symmetric wormhole solutions in Einstein-Gauss-Bonnet-dilaton gravity and showed that they are dynamically stable against spherical perturbations~\cite{kkk2011} \footnote{Recently, the instability of the present wormhole was reported in Ref. \cite{Cuyubamba:2018jdl}.}.
The Gauss-Bonnet term non-minimally coupled to a dilaton scalar field appears in the Lagrangian as the ghost-free quadratic correction in the low-energy limit of heterotic string theories~\cite{Gross}.
Although this Einstein-Gauss-Bonnet-dilaton theory is realized only in ten dimensions, their result gives courage and hope toward the construction of wormholes in our universe described by a well-motivated effective theory of gravity.
Then a natural question arises: Which is the main ingredient stabilizing the wormhole, the Gauss-Bonnet term or its dilaton coupling?

The main purpose of the article  is to clarify the effect of the Gauss-Bonnet term on the dynamical stability.
For this purpose, we will study the simplest thin-shell wormhole which is made of its tension.
While dynamical stability of thin-shell wormholes has been intensively investigated both in general relativity (Einstein gravity)~\cite{visser, poisson-visser, Reissner, ishak-lake2002, lobo-crawford2004, cylindricalTSW in Einstein gravity, Garcia+, dias-lemos2010} and in various models of modified gravity~\cite{TSWs in modified}, pure tension brane is the best set-up to analyze stability as a pure gravitational effect because such a thin-shell does not suffer from the matter instability.

In Einstein gravity, such thin-shell wormholes have been fully investigated in the previous section.
In the vacuum case, such thin-shell wormholes are stable against radial perturbations only in the hyperbolically symmetric case with negative mass in the bulk spacetime~\cite{kh2015}.

In this section, we will study the same system with the Gauss-Bonnet term but without a dilaton in the Lagrangian.
Since the Gauss-Bonnet term becomes total derivative and does not affect the field equations in four or lower dimensions in the absence of the non-minimal coupling to a dilaton, we will consider five or higher-dimensional spacetimes.
In comparison with the general relativistic case, the equation of motion for the shell is much more complicated.
For this reason, although thin-shell wormholes have been investigated in Einstein-Gauss-Bonnet gravity by many authors~\cite{shell-GB}, the stability analysis has not been completed yet even against radial perturbations.

In Sec. \ref{Stability criterion}, for investigating stability of shell wormholes, we will develop a systematic method that is applied to any gravitational theories. This method makes stability analysis simpler.

\subsection{Bulk solution}
\label{bulk}
We will study the properties of thin-shell wormholes in Einstein-Gauss-Bonnet gravity.
Such wormhole solutions are constructed by gluing two bulk solutions at a timelike hypersurface.

In this article , we consider the $d$-dimensional vacuum solution with a maximally symmetric base manifold~\cite{bdw} as the bulk solution, of which metric is given by 
\begin{align}
\D s_d^2=g_{\mu\nu}\D x^\mu \D x^\nu=-f(r)\D t^2+f(r)^{-1}\D r^2+r^2\gamma_{AB}\D z^A\D z^B,\label{BDW1}
\end{align}
where $z^A$ and $\gamma_{AB}~(A,B=2,3,\cdots,d-1)$ are the coordinates and the unit metric on the base manifold and 
\begin{align}
\label{BDW}
f(r) := k+\frac{r^2}{2\tilde{\alpha }}\left(1\mp \sqrt{1+\frac{4\tilde{\alpha }m}{r^{d-1}}+4{\tilde\alpha}{\tilde\Lambda}}\right). 
\end{align}
Here $k=1,0,-1$ is the curvature of the maximally symmetric base manifold corresponding to the spherical, planar, and hyperbolically symmetric spacetime, respectively.
$m$ is called the mass parameter.

The expression of the metric function (\ref{BDW}) shows that there are two branches of solutions corresponding to the different signs in front of the square root.
The branch with the minus sign, called the GR branch, allows the general relativistic limit $\alpha\to 0$ as
\begin{align}
\label{f-gr GR-limit}
f(r) = k-\frac{m}{r^{d-3}}-{\tilde\Lambda}r^2.
\end{align}
On the other hand, the metric in the branch with the plus sign, called the non-GR branch, diverges in this limit.
In the following section, we will consider the bulk solution only in the GR branch as a conservative choice.

The global structure of the spacetime (\ref{BDW1}) depending on the parameters has been completely classified~\cite{tm2005}.
There are two classes of curvature singularities in the spacetime.
One is the central curvature singularity at $r=0$.
Since we assume ${\tilde\alpha}>0$ and $1+4\tilde{\alpha }\tilde{\Lambda}>0$, the interior of  the square root becomes zero at some $r=r_{\rm b}(>0)$ for negative $m$.
This corresponds to another curvature singularity called the branch singularity and the metric becomes complex at $r<r_{\rm b}$.
In this case, the domain of the coordinate $r$ is $r\in (r_{\rm b},\infty)$.

The spacetime has a Killing horizon at $r=r_{\rm h}$ satisfying $f(r_{\rm h})=0$ depending on the parameters.
In order to construct static thin-shell wormholes, the bulk spacetime needs to be static.
For this reason, we consider the bulk solution (\ref{BDW1}) in the domain $r\in (r_{\rm h},\infty)$ if there is no branch singularity and $r\in (\max\{r_{\rm b},r_{\rm h}\},\infty)$ if there is a branch singularity.
We define the future (past) direction with increasing (decreasing) $t$.
We stress that since there is a generalized Birkhoff's theorem in Einstein-Gauss-Bonnet gravity \cite{Wiltshire:1985us}, the bulk spacetime is static as long as a shell moves with preserving symmetries.

\subsection{Equation of motion for a thin-shell}
A thin-shell wormhole spacetime is constructed by gluing two bulk spacetimes (\ref{BDW1}) at a timelike hypersurface $r = a$.
Here the bulk spacetimes are defined in the domain  $r \ge a(>r_{\rm h})$ and may have different parameters.
Then the junction conditions, which are the field equations (\ref{beq}) in the distributional sense, tell us the matter content on the thin-shell at $r =a$.
Finally, the equation of motion for the shell is obtained as a closed system when the equation of state for the matter is assumed.

The junction condition in Einstein-Gauss-Bonnet gravity is given by 
\begin{equation} 
\label{j-condition}
[K^i_{~j}]-\delta^i_{~j}[K]+2\alpha\Bigr(3[J^i_{~j}] -\delta^i_{~j}[J] -2 P^i_{~kjl}[K^{kl}]\Bigr)=-\kappa_d^2 S^i_{~j},
\end{equation}
$i,j=1,2,\cdots, (d-1)$ are indices for the coordinates on the timelike shell~\cite{myers1987,davis2003,GB-junction}.
In the junction conditions (\ref{j-condition}), $K^i_{~j}$ is the extrinsic curvature of the shell and $K:=h^{ij}K_{ij}$, where $h_{ij}$ is the induced metric on the shell.
Other geometrical quantities are defined by 
\begin{eqnarray}
J_{ij} &:=&{1\over 3} \left(2KK_{ik}K^{k}{}_{j}+K_{kl}K^{kl}K_{ij}-2K_{ik}K^{kl}K_{lj}-K^2 K_{ij}\right) \, , \\
P_{ikjl}&:=&{\cal R}_{ikjl}+2h_{i[l}{\cal R}_{j]k}+2h_{k[j}{\cal R}_{l]i} +{\cal R}h_{i[j}h_{l]k} \, ,
\end{eqnarray}
where ${\cal R}_{ijkl}$, ${\cal R}_{ij}$, and ${\cal R}$ are the Riemann tensor, Ricci tensor, and Ricci scalar on the shell, respectively.
$P_{ijkl}$ is the divergence-free part of the Riemann tensor ${\cal R}_{ijkl}$ satisfying $D_i P^{i}{}_{jkl}=0$, where $D_i$ is the covariant derivative on the shell.
Lastly, $S^i{}_j$ is the energy-momentum tensor on the shell, which satisfies the conservation equations $D_iS^i_{~j}=0$.

A static thin-shell wormhole is realized as a static solution for the equation of motion.
However, in general, $a$ is not constant but changes in time, representing a moving shell.
In such a case, $a$ may be written as a function of the proper time $\tau$ on the shell as $a=a(\tau)$.

Now let us derive the equation of motion for the shell.
We describe the position of the thin-shell as $r=a(\tau)$ and $t=T(\tau)$ in the spacetime (\ref{BDW1}) and assume the form of $S^i{}_j$ as
\begin{equation}
S^i{}_j = \mbox{diag} (-\rho,p,p,\cdots, p) + \mbox{diag}(-\sigma,-\sigma,-\sigma,\cdots, -\sigma)  \label{Sij}.
\end{equation}
This assumption means that the matter on the shell consists of a perfect fluid and the constant tension $\sigma$ of the shell, where $\rho$ and $p$ are the energy density and pressure of the perfect fluid.
Assuming the Z${}_2$ symmetry for the bulk spacetime, we write down the junction conditions (\ref{j-condition}) as
\begin{align} 
\frac12\kappa_d^2(\rho+\sigma)=&-\frac{(d-2)f{\dot T}}{ a}\biggl\{1+\frac{2{\tilde\alpha}}{3}\biggl( 2\frac{{\dot a}^2}{a^2}+\frac{3k}{a^2} -\frac{f}{a^2}\biggl)\biggl\}, \label{eom1}\\
-\frac12\kappa_d^2(p-\sigma)=&-\frac{a}{f{\dot T}}\biggl\{\frac{{\ddot a}}{a}+\frac{f'}{2a}+(d-3)\biggl(\frac{{\dot a}^2}{a^2}+\frac{f}{a^2}\biggl)\biggl\}  \nonumber \\
&-\frac{2{\tilde\alpha}a}{f{\dot T}}\biggl\{\frac{d-5}{3}\biggl(\frac{{\dot a}^2}{a^2}+\frac{f}{a^2}\biggl)\biggl(2\frac{{\dot a}^2}{a^2}+\frac{3k}{a^2}-\frac{f}{a^2}\biggl) \nonumber \\
&+\biggl(2\frac{{\dot a}^2}{a^2}+\frac{k}{a^2}+\frac{f}{a^2}\biggl)\frac{{\ddot a}}{a} +\frac{f'}{2a}\biggl(\frac{k}{a^2}-\frac{f}{a^2}\biggl) \biggl\},\label{eom2}
\end{align}
where $f=f(a)$.
A dot and a prime denote the derivative with respect to $\tau$ and $a$, respectively.
See 
Ref. \cite{kokubu-maeda-harada2015} for the details of derivation.
The above equations give the equation of motion for a thin-shell in Einstein-Gauss-Bonnet gravity.

In order to obtain the equation of motion in a closed system, the equation of state for the perfect fluid is required.
One possibility is the following linear equation of state with a constant $\gamma$ :
\begin{equation}
p=(\gamma-1)\rho. \label{eos}
\end{equation}
With this equation state, the energy-conservation equation on the shell $D_iS^i_{~j}=0$, written as
\begin{equation}
{\dot \rho}=-(d-2)(p+\rho)\frac{\dot a}{a}, \label{em-cons}
\end{equation}
is integrated to give
\begin{equation}
\rho=\frac{\rho_0}{a^{(d-2)\gamma}}, \label{energy}
\end{equation}
where $\rho_0$ is a constant.

\subsection{Effective potential for the shell}
The dynamics of the shell governed by Eqs.~(\ref{eom1}) and (\ref{eom2}) with the equation of state (\ref{eos}) can be discussed as a one-dimensional potential problem.
Then the shape of the effective potential $V(a)$ for the shell determines the stability of static configurations, namely the static wormholes.

Let us derive the effective potential $V(a)$.
Squaring Eq.~(\ref{eom1}) and using $u^\mu u_\mu=-1$, we obtain
\begin{align} 
\Omega(a)^2=&\biggl(\frac{f}{a^2}+\frac{{\dot a}^2}{a^2}\biggl)\biggl\{1 +\frac{2}{3}{\tilde\alpha}\biggl( 2\frac{{\dot a}^2}{a^2}+\frac{3k}{a^2}-\frac{f}{a^2} \biggl)\biggl\}^2, \label{eom3}
\end{align}
where
\begin{align}
\Omega(a)^2:=&\frac{\kappa_d^4(\rho(a)+\sigma)^2}{4(d-2)^2}. \label{Omega01}
\end{align}
This is a cubic function for ${\dot a}^2$.
The position of the throat $a=a_0$ for a static wormhole is obtained by solving the following algebraic equation for $a_0$:  
\begin{align} 
\Omega_0^2=\frac{f_0}{a_0^2}\biggl\{1 +\frac{2}{3}{\tilde\alpha}\biggl(\frac{3k}{a_0^2}-\frac{f_0}{a_0^2} \biggl)\biggl\}^2, \label{omegastatic}
\end{align}
where $f_0:=f(a_0)$ and $\Omega_0^2:=\Omega(a_0)^2$.

For convenience, we define
\begin{align}
A(r):=&1+\frac{4\tilde{\alpha }m}{r^{d-1}}+4{\tilde\alpha}{\tilde\Lambda}
\end{align}
with which the metric function (\ref{BDW}) in the GR branch is written as
\begin{align}
f(r) =& k+\frac{r^2}{2\tilde{\alpha }}\left(1-\sqrt{A(r)}\right). \label{metric}
\end{align}
$A>0$ is required for the real metric and the absence of branch singularity. 
In the GR branch, because of the existence of the square root in Eq. (\ref{metric}), the following inequality holds:
\begin{align}
r^2+2{\tilde\alpha}k-2{\tilde\alpha}f(r)>0, \label{GR branch ineq}
\end{align} 
which will be used later.

Actually, Eq.~(\ref{eom3}) admits only a single real solution for ${\dot a}^2$:
\begin{align}
{\dot a}^2=-V(a), \label{conservation law}
\end{align}
which has the form of the one-dimensional potential problem.
The effective potential $V(a)$ is defined by 
\begin{align}
V(a):=f(a)-J(a)a^2, \label{potential}
\end{align}
where $J(a)$ is defined by
\begin{align}
J(a):=&\frac{\left(B(a)-A(a)^{1/2}\right)^2}{4{\tilde\alpha}B(a)}, \label{j}\\
B(a):=&\biggl\{18{\tilde\alpha}\Omega(a)^2+A(a)^{3/2}+6\sqrt{{\tilde\alpha}\Omega(a)^2(9{\tilde\alpha}\Omega(a)^2+A(a)^{3/2})}\biggl\}^{1/3}.
\end{align}
One can see $B>A^{1/2}$.
$\Omega^2$ can be expressed in terms of $A$ and $B$ as
\begin{align}
\Omega^2=\frac{(B^3-A^{3/2})^2}{36{\tilde\alpha}B^3}. \label{Omega02}
\end{align}

\subsection{Negative energy density of the shell} 
\label{Energy density}
We show that the energy density on the shell $\rho+\sigma$ must be negative for static wormholes.
The condition $\rho+\sigma\ge 0$ and Eq. (\ref{eom1}) with $a=a_0(>0)$ yields $a_0^2 \le -4{\tilde\alpha}k/(2+\sqrt{A_0})$, where $A_0:=A(a_0)$.
Clearly, this is not satisfied for $k=1,0$ under the assumption ${\tilde\alpha}>0$.
For $k=-1$. Eq.~(\ref{eom1}) gives $a_0^2 \le 4{\tilde\alpha}/(2+\sqrt{A_0}) <2{\tilde\alpha}$ and this is not satisfied because there is a constraint $a_0^2>2{\tilde\alpha}$ for the throat radius in the GR branch, which can be shown from the combination of Eq. (\ref{GR branch ineq}) and $f(a_0)>0$.
Now we have shown that the energy density on the shell is negative in the physical set up considered in this article.  Hence, the weak energy condition is violated in the GR-branch with $\alpha>0$. However, the negative-tension brane still satisfies the null energy condition.

Hereafter we will consider the case without a perfect fluid on the shell ($\rho=p=0$) and assume $\sigma<0$.
The resulting static thin-shell wormholes are made of the pure (negative) tension $\sigma$ and satisfy the null energy condition.
This simplest set up is preferred by the minimal violation of the energy conditions.
The energy-momentum tensor for the negative tension is equivalent to a negative cosmological constant. Such a matter corresponds to an inverted harmonic oscillator in classical mechanics.
A technical advantage in this set up is the constancy of $\Omega^2$.

\subsection{Static solutions} 
Here we summarize the existence conditions for a static shell located at $a=a_0$.
Equation~(\ref{conservation law}) is interpreted as the conservation law of mechanical energy for the shell. 
By differentiating Eq.~(\ref{conservation law}) with respect to $\tau$, we obtain the equation of motion for the shell as
\begin{eqnarray}
\ddot a=-\frac{1}{2}V^\prime (a). \label{EOM}
\end{eqnarray}
From Eqs.~(\ref{conservation law}) and (\ref{EOM}), $a_0$ is determined algebraically by $V(a_0)=0$ and $V^\prime(a_0)=0$.
In addition, $a_0$ must satisfy $A(a_0)>0$ and $f(a_0)>0$. 
The latter condition $f(a_0)>0$ is called the horizon-avoidance condition in the previous section and Ref.~\cite{barcelo&visser}, which simply means that the position of the throat is located in the static region of the spacetime.
Actually, this condition is always satisfied because we have $V(a_0)=0$ and Eq.~(\ref{potential}) implies $f(a)>V(a)$.

The location of the static wormhole throat $a=a_0$ is determined by the following algebraic equation obtained by eliminating $\sigma$ from Eqs. (\ref{eom1}) and (\ref{eom2}):
\begin{align}
\biggl(\frac{f_0}{a_0}-\frac{f_0^\prime}{2}\biggl)\biggl(1+2{\tilde\alpha}\frac{k-f_0}{a_0^2}\biggl)+\frac{8 {\tilde\alpha}kf_0}{a_0^3}=0, \label{static01}
\end{align}
where $f_0':=f'(a_0)$.
In the limit of $\alpha \rightarrow 0$, Eq.~(\ref{static01}) reduces to the corresponding equation in Einstein gravity~\cite{kh2015}.
The explicit form of Eq.~(\ref{static01}) is 
\begin{align}
ka_0^2\sqrt{A_0}= 2ka_0^2-\frac{(d-1)m}{2a_0^{d-5}}+4{\tilde\alpha}k^2. \label{static02}
\end{align}

\subsection{Stability criterion} \label{Stability criterion} 
As explained at the beginning of this section, the sign of $V''(a_0)$ determines stability of a static thin-shell wormhole.
In this subsection, we will derive $V''(a_0)$ in a convenient form to prove the (in)stability.

\subsubsection{Einstein gravity} 
First let us consider Einstein gravity as a simple lesson.  
In the general relativistic limit $\alpha\to 0$, Eq.~(\ref{eom3}) reduces to 
\begin{align}
\Omega^2=&\frac{f(a)}{a^2}+\frac{{\dot a}^2}{a^2}. \label{eom3-gr0}
\end{align}
By solving Eq. (\ref{eom3-gr0}) for ${\dot a}^2$, we define a potential $V(a)$ of the conservation law of the one-dimensional potential problem. Then we directly calculate the second derivative of $V(a)$. However, without such direct calculations, in principle we can derive the form of $V''(a_0)$ by operating a systematic method below, which can be applied also in more general theories of gravity. 

Suppose we get a master equation as ${\dot a}^2+V(a)=0$.
By this master equation, ${\dot a}^2$ in Eq.~(\ref{eom3-gr0}) is replaced by $-V(a)$ to give
\begin{align}
\Omega^2=&\frac{f(a)-V(a)}{a^2}. \label{eom3-gr}
\end{align}
Differentiating this equation twice, we obtain
\begin{align}
0=&a(f'-V')-2(f-V), \label{eom3-gr2}\\
0=&a(f''-V'')-(f'-V'), \label{eom3-gr3}
\end{align}
where $\Omega'=0$ for the pure tension brane.

In Einstein gravity, the metric function is 
\begin{align}
f(a)=&k-\frac{m}{a^{d-3}}-{\tilde\Lambda}a^2,\label{f-gr}
\end{align}
which satisfies
\begin{align}
f'(a)=&\frac{(d-3)(k-f)-{\tilde\Lambda}(d-1)a^2}{a},\label{df-gr}\\
f''(a)=&\frac{{\tilde\Lambda}(d-1)(d-4)a^2-(k-f)(d-2)(d-3)}{a^2}. \label{ddf-gr}
\end{align}

Substituting Eq.~(\ref{df-gr}) into Eq.~(\ref{eom3-gr2}) and evaluating it at $a=a_0$ satisfying $V(a_0)=V'(a_0)=0$, we obtain
\begin{align}
f_{{\rm E}0}(:=f(a_0))=\frac{d-3}{d-1}k-{\tilde\Lambda}a_0^2. \label{key-gr1}
\end{align}
Combining this with Eq.~(\ref{f-gr}), we obtain the algebraic equation to determine $a_0$:
\begin{align}
\frac{2k}{d-1}=&\frac{m}{a_0^{d-3}}. \label{key-gr2}
\end{align}
 For $k=0$, Eq.~(\ref{key-gr2}) requires $m=0$ and $a_0$ is totally undetermined.
For $k=1(-1)$, Eq.~(\ref{key-gr2}) requires $m>(<)0$ and the throat radius $a_0$ is given by
\begin{align}
a_0=\biggl(\frac{(d-1)m}{2k}\biggl)^{1/(d-3)}. \label{solution-GR}
\end{align}
As seen in Eq.~(\ref{solution-GR}), $\Lambda$ does not contribute to the size of the wormhole throat. However, it appears in the horizon-avoidance condition $f_0>0$.
Equation~(\ref{key-gr1}) shows that $f_0>0$ requires $\Lambda<0$ in the case of $k=0,-1$.
In the case of $\Lambda=0$,  $f_0>0$ is satisfied only for $k=1$.
In the case of $\Lambda>0$ and $k=1$, $f_0>0$ gives a constraint $a_0<a_{\rm c}^{\rm(GR)}$ on the size of the wormhole throat, where 
\begin{align}
a_{\rm c}^{\rm(GR)}:=&\biggl(\frac{(d-3)k}{(d-1){\tilde\Lambda}}\biggl)^{1/2}. \label{a-c-gr}
\end{align}
On the other hand, in the case of $\Lambda<0$ and $k=-1$, $f_0>0$ gives  $a_0>a_{\rm c}^{\rm(GR)}$.
Combining this inequality with Eq.~(\ref{solution-GR}), we obtain the range of the mass parameter admitting static wormhole solutions; $0<m<m_{\rm c}^{\rm(GR)}$ for $k=1$ with $\Lambda>0$ and $m<m_{\rm c}^{\rm(GR)}(<0)$ for $k=-1$ with $\Lambda<0$, where
\begin{align}
m_{\rm c}^{\rm(GR)}:=&\frac{2k}{d-1}\biggl(\frac{(d-3)k}{(d-1){\tilde\Lambda}}\biggl)^{(d-3)/2}. \label{m-c-gr}
\end{align}

In Einstein gravity, a simple criterion for the stability of static solutions is available. 
Substituting Eqs.~(\ref{df-gr}) and (\ref{ddf-gr}) into Eq.~(\ref{eom3-gr3}), evaluating them at $a=a_0$, we obtain
\begin{align}
V''(a_0)=&-\frac{(d-1)(d-3)m}{a_0^{d-1}} 
=-\frac{2(d-3)k}{a_0^2}, \label{ddV-gr-final}
\end{align}
where we used Eqs.~(\ref{f-gr}) and (\ref{key-gr2}).
This simple expression clearly shows that the wormhole is stable only for $k=-1$ with $m<0$~\cite{kh2015}.
Existence and stability of static thin-shell wormholes in Einstein gravity are summarized in Table~\ref{table:results-GR}.
\begin{table*}[htb]
\begin{center}
\caption{The existence and stability of Z${}_2$ symmetric static thin-shell wormholes made of pure negative tension in Einstein gravity, where $a_{\rm c}^{\rm(GR)}$ and $m_{\rm c}^{\rm(GR)}$ are defined by Eqs.~(\ref{a-c-gr}) and (\ref{m-c-gr}), respectively.  }
\label{table:results-GR}
\begin{tabular}{|c|c|c|c|c|}\hline
& & Existence  & Possible range of $a_0$ & Stability \\ \hline
$k=1$& $\Lambda > 0$ &  $0<m<m_{\rm c}^{\rm(GR)}$  & $0<a_0<a_{\rm c}^{\rm(GR)}$  & Unstable \\ \cline{2-5}
 & $\Lambda\le 0$ & $m>0$   & $a_0>0$  & Unstable \\ \hline
$k=0$ & $\Lambda \ge 0$ & None  & -- & -- \\ \cline{2-5}
& $\Lambda< 0$ & $m=0$  & $a_0>0$ & Marginally stable \\ \hline
$k=-1$& $\Lambda \ge 0$ &  None  & --  & -- \\ \cline{2-5}
 &   $\Lambda< 0$ & $m<m_{\rm c}^{\rm(GR)}(<0)$ & $a_0>a_{\rm c}^{\rm(GR)}$ & Stable \\ \hline
\end{tabular}
\end{center}
\end{table*}

\subsubsection{Einstein-Gauss-Bonnet gravity} 
Although it is more complicated, we can play this game in Einstein-Gauss-Bonnet gravity in a similar manner.
Replacing ${\dot a}^2$ by $-V(a)$ in the master equation (\ref{eom3}), we obtain
\begin{align}
\Omega^2=&\frac{f(a)-V(a)}{a^2}\biggl\{1 +\frac{2{\tilde\alpha}(-2V(a)+3k-f(a))}{3a^2}\biggl\}^2. \label{eom3-3}
\end{align}

In Einstein-Gauss-Bonnet gravity, the metric function is 
\begin{align}
f(a)=&k+\frac{a^2}{2{\tilde\alpha}}\biggl(1-\sqrt{1+4{\tilde\alpha}{\tilde\Lambda}+\frac{4{\tilde\alpha}m}{a^{d-1}}}\biggl),\label{f-GB}
\end{align}
which satisfies
\begin{align}
f'(a)=&\frac{(d-5){\tilde\alpha}(k-f)^2+(d-3)a^2(k-f)-{\tilde\Lambda}(d-1)a^4}{a\{a^2+2{\tilde\alpha}(k-f)\}},\label{df-GB}\\
f''(a)=&\frac{L(a)}{a^2\{a^2+2{\tilde\alpha}(k-f)\}^3},\label{ddf-GB}
\end{align}
where
\begin{align}
L(a):=&2(d-1)^2{\tilde\alpha}{\tilde\Lambda}^2a^8-{\tilde\Lambda}a^4(d-1)\biggl\{12{\tilde\alpha}^2(k-f)^2+12{\tilde\alpha}a^2(k-f)-(d-4)a^4\biggl\} \nonumber \\
&-(k-f)\biggl\{2(d-3)(d-5){\tilde\alpha}^3(k-f)^3+4(d^2-8d+13){\tilde\alpha}^2a^2(k-f)^2 \nonumber \\
&+3(d-2)(d-5){\tilde\alpha}a^4(k-f)+(d-2)(d-3)a^6\biggl\}.
\end{align}
Equation~(\ref{f-GB}) gives
\begin{align}
m=a^{d-3}\biggl\{-{\tilde\Lambda}a^2+(k-f(a))+{\tilde\alpha}a^{-2}(k-f(a))^2\biggl\}.\label{mf-GB}
\end{align}

Differentiating (\ref{eom3-3}) and evaluating it at $a=a_0$, we obtain
\begin{align}
f_0^2=&\frac{\{(d-1)a_0^2+2k(d+1){\tilde\alpha}\}f_0+(d-1){\tilde\Lambda}a_0^4-k\{(d-3)a_0^2+(d-5){\tilde\alpha}k\}}{(d-1){\tilde\alpha}}. \label{f^2-GB}
\end{align}
where we used Eq.~(\ref{df-GB}).
This equation reduces to Eq.~(\ref{key-gr1}) for $\alpha\to 0$.
Equation~(\ref{f^2-GB}) will be used to express $f_0^p~(p=2,3,4,\cdots)$ in terms of $f_0$.

Differentiating Eq.~(\ref{eom3-3}) twice and using Eqs.~(\ref{df-GB}) and (\ref{ddf-GB}), we finally obtain $V''(a_0)$ in a rather compact form:
\begin{align}
V''(a_0)=&-\frac{2kP(a_0)}{a_0^2(a_0^2+2k{\tilde\alpha}+2{\tilde\alpha}f_0)(a_0^2+2k{\tilde\alpha}-2{\tilde\alpha}f_0)}, \label{ddV-gb} \\
P(a_0):= &~ 4{\tilde\alpha}^2f_0\biggl\{6k-(d-3)f_0\biggl\}+(a_0^2+2k{\tilde\alpha})\biggl\{(d-3)a_0^2+2(d-5)k{\tilde\alpha}\biggl\}, \label{defP}
\end{align}
where we have eliminated ${\tilde\Lambda}$ by using Eq.~(\ref{f^2-GB}).
This expression reduces to Eq.~(\ref{ddV-gr-final}) for $\alpha\to 0$.
Because of Eq.~(\ref{GR branch ineq}), the denominator in the expression of $V''(a_0)$ is positive and, therefore, the sign of the function $P(a_0)$ determines the stability of the shell.

\subsection{Effect of the Gauss-Bonnet term on the stability}
Before moving onto the full-order analysis, let us clarify how the Gauss-Bonnet term affects the stability of thin-shell wormholes in the situation where ${\tilde\alpha}$ is small.

In the general relativistic limit ${\tilde\alpha} \to 0$ with $k=\pm1$,  Eq.~(\ref{static02}) gives
\begin{align}
a_0^{d-3}=\frac{(d-1)mk}{2}=:a_{{\rm E}}^{d-3}.
\end{align}
This is the static solution in Einstein gravity which requires $mk>0$~\cite{kh2015}. 
Now we obtain the static solution for $\epsilon:={\tilde \alpha}/a_E^2\ll 1$ in a perturbative method. We expand $a_0$ in a power series of $\epsilon$ :
\begin{align}
a_0=a_{{\rm E}}+a_{(1)}\epsilon+a_{(2)}\epsilon^2+\dots. \label{taylor}
\end{align}
Substituting this expression into Eq. (\ref{static02}) and expanding it in a series of $\epsilon$, we obtain
\begin{align}
a_{(1)}=\frac{2{\tilde\Lambda}(d-1)a_{{\rm E}}^2-4(d-2)k}{(d-1)(d-3)}a_{{\rm E}}. \label{a(1)}
\end{align}
This allows us to derive the expansion of Eq. (\ref{ddV-gb}):
\begin{align}
V^{\prime\prime}(a_0)\simeq -\frac{2(d-3)k}{a_{{\rm E}}^2}-\frac{8k\epsilon}{a_{{\rm E}}^2}f_{{\rm E}0}. \label{ddV-GB-linear-2}
\end{align}
The first term of Eq.~(\ref{ddV-GB-linear-2}) coincides with Eq.~(\ref{ddV-gr-final}) and
 the second term is negative because of Eq.~(\ref{key-gr1}). 
Hence, we arrive at a simple conclusion about the effect of the Gauss-Bonnet term for small ${\tilde\alpha}$; it destabilizes wormholes in the spherically symmetric case ($k=1$), while it stabilizes in the hyperbolically symmetric case ($k=-1$).

\subsection{Stability analysis}
Here, we will prove (in)stability of the static thin-shell wormhole in the framework of Einstein-Gauss-Bonnet gravity.
We focus on stability analysis for $k=1$ as an educational example.

\begin{figure}[htbp]
\begin{center}
\includegraphics[width=0.6\linewidth]{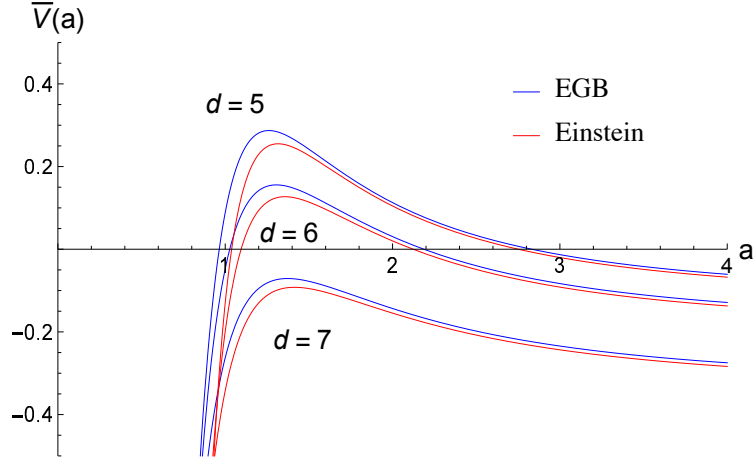}
\caption{\label{fig:EGBGR} The potential $\bar V (a)$ for $d=5,6,7$ in
 Einstein and Einstein-Gauss-Bonnet (EGB) 
gravity with $k=1$, $\alpha=0.02$, $m=1$, $\Lambda=1$ and $\sigma=-0.1$.}
\end{center}
\end{figure}

In this article, we do not clarify the parameter region with positive $m$ admitting static wormhole solutions because they are all dynamically unstable in any case.
In Fig.~\ref{fig:EGBGR}, we plot the profile of $\bar V (a)$ with $k=1$ and $m>0$, in which there is a local maximum. This implies that the corresponding static solution is unstable.
Indeed, we can prove that this case is unstable.
For $k=1$, the positivity of $P(a_0)$ in Eq.~(\ref{ddV-gb}) means instability of the static wormhole.
Using the inequality (\ref{GR branch ineq}), we evaluate the lower bound of $P(a_0)$ as
\begin{align}
P(a_0)=&4{\tilde\alpha}^2f_0\biggl\{6-(d-3)f_0\biggl\}+(a_0^2+2{\tilde\alpha})\biggl\{(d-3)a_0^2+2(d-5){\tilde\alpha}\biggl\} \nonumber \\
> &4{\tilde\alpha}^2f_0\biggl\{6-(d-3)f_0\biggl\}+2{\tilde\alpha}f_0\biggl\{(d-3)a_0^2+2(d-5){\tilde\alpha}\biggl\} \nonumber \\
= &2{\tilde\alpha}f_0\biggl\{2(d-3){\tilde\alpha}\biggl(\frac{a_0^2}{2{\tilde\alpha}}-f_0\biggl)+2(d+1){\tilde\alpha}\biggl\}  \nonumber \\
>&2{\tilde\alpha}f_0\biggl\{-2(d-3){\tilde\alpha}+2(d+1){\tilde\alpha}\biggl\}=16{\tilde\alpha}^2f_0>0.
 \label{eval-P}
\end{align}
Therefore, the wormhole is dynamically unstable.
We note that a similar analysis for $k=0,1$ can reveal the instability of thin-shell wormholes which are made of a dust fluid. (See Ref. \cite{kokubu-maeda-harada2015}.)

We have focused on the analysis with $k=1$ and $m>0$. The rest of combinations of $k$ and $m$ has also been completely analyzed. See Ref. \cite{kokubu-maeda-harada2015} for the detail. 
All the results obtained in Ref. \cite{kokubu-maeda-harada2015} for all combinations of $k$ and $M$ are summarized in Table~\ref{table:results}.

\begin{table*}[htb]
\begin{center}
\caption{The existence and stability of Z${}_2$ symmetric static thin-shell wormholes made of pure negative tension in the GR branch with ${\tilde\alpha}>0$ and $1+4{\tilde\alpha}{\tilde\Lambda}>0$. "S", "M", and "U" stand for "Stable", "Marginally stable", and "Unstable", respectively }
\label{table:results}
\begin{tabular}{|c|c|c|c|}\hline
& & Static solutions exist?  & Stability \\ \hline
$k=1$& $m>0$ & Yes  & U \\ \cline{2-4}
& $m \le 0$ & No & -- \\ \hline
$k=0$ & $m=0$ & $\Lambda\ge 0$: No   & -- \\ \cline{3-4}
&          &  $\Lambda < 0$: Yes  & M  \\ \cline{2-4}
& $m \ne 0$ & No  & -- \\ \hline
$k=-1$& $m \ge 0$ &  No  & --  \\ \cline{2-4}
 & $m<0$ &  ${\Lambda} \ge 0$ : No  & --  \\ \cline{3-4}
&          &  $-(2d-5)/(2d-1) < 4{\tilde\alpha}{\tilde\Lambda} < 0$: Yes  & S  \\ \cline{3-4}
&          &  $4{\tilde\alpha}{\tilde\Lambda}=-(2d-5)/(2d-1)$: Yes  & S or M \\ \cline{3-4}
 &         & $-1< 4{\tilde\alpha}{\tilde\Lambda} < -(2d-5)/(2d-1)$ with $d=5$: Yes & S or M \\ \cline{3-4}
 &         & $-1< 4{\tilde\alpha}{\tilde\Lambda} < -(2d-5)/(2d-1)$ with $d\ge 6$: Yes & S, M, or U \\ \hline
\end{tabular}
\end{center}
\end{table*}

\newpage
\section{Conclusion}\label{Discussions and conclusions}
\subsection{In Einstein gravity}
We developed the thin-shell formalism for $d$ dimensional spacetimes,
which is more general than Dias and Lemos formalism~\cite{dias-lemos2010}. We investigated spherically, planar
and hyperbolically symmetric wormholes 
with a pure negative-tension brane and found and classified Z${}_2$ symmetric static solutions which are stable
against radial perturbations. 
We summarized the results in Tables \ref{table d=3}, \ref{table k=+1}, \ref{table k=-1} and \ref{table k=0}.
We found that in most cases
charge is needed to keep the static throat radius positive and that a
negative cosmological constant tends to decrease the radius of the black
hole horizon and then to achieve the horizon avoidance. So, the
combination of an electric
charge and a negative cosmological constant makes it easier to construct
stable wormholes. However, a negative cosmological constant is
unnecessary in a certain situation of $k=+1$ and $M>0$ and charge is
unnecessary in a certain situation of $k=-1$ and $M<0$.

We would note that the existence and stability of
negative-tension branes as thin-shell wormholes crucially depend on the curvature of 
the maximally symmetric $(d-2)$ dimensional manifolds. 
On the other hand, they do not qualitatively
but only quantitatively depend on the number of spacetime dimensions.

We considered only radial stability for thin-shell wormholes here. In a realistic situation, a wormhole in the universe would be suffered by gravitational waves that are produced by a particle falling into the wormhole, or, incidental waves produced from gravitational collapses of nearby stars, etc. In such situations we must consider non-radial gravitational perturbations for wormholes. So far, there is no stability analysis for gravitational perturbations of thin-shell wormholes. 
A linear stability analysis on a specific-perturbation mode (so-called axial perturbations) for a wormhole that corresponds to Morris-Thorne type has been investigated \cite{bronnikov+2013}. In \cite{bronnikov+2013}, the matter distribution is continuous not as a shell. Since matters of thin-shell wormholes are localized on their throat as a shell, the perturbation analysis for thin-shell wormholes is not like that of Morris-Thorne type.
To treat the non-spherically perturbed shell, we can employ the perturbed junction condition developed by Gerlach and Sengupta \cite{gerlach-sengupta}. By using the perturbed junction condition, Kodama {\it et al.} investigated whether a domain wall emits gravitational waves or not. In their work, a domain wall is constructed by pasting a couple of Minkowski spacetimes with a negative-tension brane localized at their boundary \cite{ishibashi-kodama1994}. Since their situation is similar to that of a thin-shell wormhole, we expect their study can be extended to the case of shell wormholes.

\subsection{In Einstein-Gauss-Bonnet gravity}
$d(\ge 5)$-dimensional static thin-shell wormholes with the Z${}_2$ symmetry have been investigated in the spherically ($k=1$), planar ($k=0$), or hyperbolically ($k=-1$) symmetric spacetime 
in Einstein-Gauss-Bonnet gravity.
For our primary motivation to reveal the effect of the Gauss-Bonnet term on the static configuration and dynamical stability of a wormhole, we have studied the stability against linear perturbations preserving symmetries in the simplest set up where the thin-shell is made of pure negative tension, which satisfies the null energy condition. 

In this system, the dynamics of the shell can be treated as a one-dimensional potential problem characterized by a mass parameter $m$ in the vacuum bulk spacetime for a given value of $d$, $k$, the cosmological constant $\Lambda$, and the Gauss-Bonnet coupling constant $\alpha$.
We have studied solutions which admit the general relativistic limit $\alpha\to 0$ and considered a very conservative region in the parameter space.
The shape of the effective potential for the shell dynamics clarifies possible static configurations of a wormhole and their dynamical stability.

As seen in Tables~\ref{table:results-GR} and \ref{table:results}, the results with and without the Gauss-Bonnet term are similar in many cases.
For $k=1$, static wormholes require $m>0$ and they are dynamically unstable.
For $k=0$, static wormholes require $m=0$ and $\Lambda<0$ and they are marginally stable.
For $k=-1$, $m<0$ and $\Lambda<0$ are required for static wormholes.

We have clarified the effect of the Gauss-Bonnet term on the stability in a perturbative method by expanding the equation in a power series of $\tilde \alpha$.
We have shown that, for ${\tilde \alpha}/a_E^2\ll 1$, the Gauss-Bonnet term tends to destabilize spherically symmetric thin-shell wormholes ($k=1$), while it stabilizes hyperbolically symmetric wormholes ($k=-1$). 
For planar symmetric wormholes ($k=0$), the Gauss-Bonnet term does not affect their stability and they are marginally stable, same as in Einstein gravity. 
However, we have observed that the non-perturbative effect is quite non-trivial. 

As the effect of the Gauss-Bonnet term on the existence and stability of static wormholes has been revealed in the article, the effect of its dilaton coupling is now of great interest.
Unfortunately, in the presence of a dilaton, exact bulk solutions are not
available to construct thin-shell wormholes.
Moreover, Birkhoff's theorem does not apply to
the dilatonic system because of the additional dynamical degree of
freedom in the dilaton.
Nevertheless, this is a promising direction of research leading to understand the result in~\cite{kkk2011}.

\section*{Author Contributions:} Conceptualization, T.K. and T.H.; Methodology, T.K. and T.H; Validation, T.K. and T.H.; Formal Analysis, T.K. and T.H; Investigation, T.K. and T.H; Writing – Original Draft Preparation, T.K.; Writing – Review \& Editing, T.K. and T.H; Supervision, T.H.

\section*{Funding:} This work was partially supported by JSPS KAKENHI Grant No. JP19K03876 (T. H.) 
and the National Natural Science Foundation of China under Grant No.~11690034 (T. K.).

\section*{Conflicts of Interest:} The authors declare no conflict of interest.

\end{document}